\documentclass[aps,prb,twocolumn,superscriptaddress,showpacs]{revtex4-1}

\usepackage{graphicx}
\usepackage{amsmath}
\usepackage{graphics}
\usepackage{exscale}
\usepackage{epsfig}

\renewcommand{\epsilon}{\varepsilon}

\newcommand{\integral}[3]{\!\int\limits_{#2}^{#3}\!\!{\rm d}#1\;}

\newcommand{\e}{\mathrm e}

\newcommand{\hc}{\mathrm{h.c.}}

\begin{document}

\title{Conditions for observing emergent SU(4) symmetry in a double quantum dot}
\author{Yunori Nishikawa}
\affiliation{Dept. of Physics, Osaka City University, Sumiyoshi-ku,
Osaka 558-8585 Japan.}
\author{ Oliver J. Curtin}
\affiliation{Dept. of Physics, Imperial College, London SW7 2AZ, UK.}
\author{ Alex C. Hewson}
\affiliation{Dept. of Mathematics, Imperial College, London SW7 2AZ, UK.}
\author{Daniel J.G. Crow}
\affiliation{Dept. of Mathematics, Imperial College, London SW7 2AZ, UK.}
\author{Johannes Bauer}
\affiliation{Dept. of Mathematics, Imperial College, London SW7 2AZ, UK.}

\date{\today} 
\begin{abstract}

We analyze conditions for the observation of a low energy
SU(4) fixed point in capacitively coupled  quantum dots. One problem, due to
dots with different couplings to their baths, has been considered by Tosi, Roura-Bas and Aligia (2015).
They showed how symmetry can be effectively restored via the adjustment of individual gates voltages, but they make the assumption of
infinite on-dot and inter-dot interaction strengths.
 A related  problem 
is the  difference in the magnitudes between the on-dot and interdot strengths
for capacitively coupled quantum dots.
 Here we examine  both factors,
based on a two site Anderson model, using the numerical renormalization group
to calculate the local spectral densities on the dots and the renormalized 
parameters that specify the low energy fixed point.  Our results support the conclusions
of Tosi et al. that low energy SU(4) symmetry can be restored, but asymptotically
 achieved only if the inter-dot
interaction $U_{12}$ is greater than or of the order of the band width of
the coupled conduction bath $D$, which might be difficult to achieve
experimentally. By comparing the SU(4) Kondo results for a total dot occupation $n_{\rm tot}=1$
and $n_{\rm tot}=2$ 
we conclude that the temperature dependence of the conductance is  largely determined by
the constraints of the Friedel sum rule rather than the SU(4) symmetry and suggest that
an initial increase of the conductance with temperature is a distinguishing characteristic
feature 
of an $n_{\rm tot}=1$ universal SU(4) fixed point.

\end{abstract}

\maketitle

\section{Introduction}

Quantum dots have proved to be ideal systems for studying the low energy
behavior of strongly correlated local systems as described by single impurity
Anderson and Kondo models. This is because the energy levels on the dots
and their connections to a host electron bath can be controlled and manipulated
by applied gate voltages. This has enabled the  Kondo effect, arising from
the SU(2) spin degeneracy in a single quantum dot, to be probed experimentally.
 Measurements of the current flow through
the dot, due to an applied bias voltage, have revealed in detail the many-body
low temperature induced resonance in the local density of states in the Kondo
regime\cite{GSMAMK98,COK98,WFFETK00,KAGGKS04,AGKK05}. Interest has naturally
 moved on to the observation of other
types of strong correlation states.\par

There have  been several theoretical papers dealing with the possibility
of observing an SU(4) Kondo state in a capacitively coupled double quantum dot
 \cite{BZHHD03,MGL06,TRA13,NHCB13,FMZ14,BGS14,TRA15}.
 In this arrangement an SU(2) pseudospin symmetry is introduced in addition
to the SU(2) spin symmetry by using two identical quantum dots with a
total occupation number for the double dot maintained such that $n_{\rm tot}=1$. 
 The occupation of dot 1 then corresponds to an 'up' 
pseudospin and the occupation number for the second dot 2 as 
pseudospin 'down'. One motivation for such  an arrangement is that
it allows one to measure 'spin' polarized currents 
 without the need to introduce a local magnetic field \cite{BFM12}.

Recent experimental work\cite{HHWK08,AKRCKSOG13} in which  the electron
transport through the individual dots has been measured for such a system,
has revealed directly the effects 
of the pseudospin degrees of freedom,
and provided some support for the interpretation as 
arising from an SU(4) fixed point \cite{KAW14}.

In our earlier work \cite{NHCB13} on this topic
we used the numerical renormalization group (NRG)  to calculate the renormalized parameters
which specify the effective Hamiltonian which determines the low energy behavior of the system.
This led us to the conclusion that it would be difficult to observe complete SU(4) low energy
fixed point behavior due to the  smaller inter-dot interaction compared with the
on-dot term. Here we expand upon that work to get some estimates as to whether  
 an SU(4) fixed point can be realized, given experimentally accessible ranges for the inter-dot and on-dot
interactions, $U_{12}$ and $U$, respectively. We can check, in the case that it is not completely 
realized, how close the low temperature
behavior is to such a fixed point.   
 By comparing the full spectral density on the dots with that derived
in terms of the low energy quasiparticles we can also test the range of the low
energy effective theory. There can  also be a potential problem, considered by Tosi, Roura-Bas
and Aligia \cite{TRA15},
 arising from a lack of symmetry due to from different couplings
between the baths and their respective dots. They showed, however, that when both $U_{12}$ and $U$ are taken as infinitely large,
 this symmetry can be restored by appropriately adjusting the gate voltages to each dot.
 Here we test whether or not their conclusion holds for finite strength interactions $U_{12}$ and $U$.\par

 To get a clearer understanding of the physics in the strong correlation regime we compare
the SU(4) system with $n_{\rm tot}=1$, which is due to a combination of spin and pseudospin,
with that for  $n_{\rm tot}=2$, which is due to spin alone. We estimate and compare the leading temperature corrections
to the zero bias conductances through the individual dots in the two cases. We conclude that temperature dependence of the conductances 
reflect the general features of the quasiparticle spectra rather than any strict symmetry conditions
at the low energy fixed point. 
    
\section{Double-Dot Model}

The capacitively coupled quantum dot system can be described by a two site Anderson
model of the form,    
\begin{equation}
  H=\sum_{i=1,2}(H_{i}+H_{{\rm bath},i}+ H_{{\rm
    c},i}) + H_{12},
\label{h1}
\end{equation}
where $H_{i}$ describes the individual dots, $i=1,2$, $H_{{\rm
    bath},i}$ the baths to which the dots are individually coupled by
a term $H_{{\rm   c},i}$, and $H_{12}$  is the 
interaction between the dots. 
A reasonable approximation is to take the  baths, two for each dot, to be
described by a free electron model,
\begin{equation}
H_{{\rm bath},i}=\sum_{{\bf k},\alpha,{\sigma}}\epsilon_{\bf k} c^\dagger_{{\bf k}, i,\alpha,\sigma}
c^{}_{{\bf k},i,\alpha,\sigma} 
\end{equation}
where $\alpha=s,d$ (source, drain) and $\epsilon_{\bf k}$ is an energy level in a bath,
taken to be independent of $\alpha$, $i$ and $\sigma$.\par

The Hamiltonian describing the dots $H_{i}$ is taken in the form,
\begin{equation}
H_{i}= \sum_{{\sigma}}\epsilon_{i,\sigma}d^\dagger_{i,\sigma}
 d^{}_{i,\sigma} 
 +  U_i n_{i,\uparrow}n_{i,\downarrow},
\end{equation}
where $\epsilon_{i,\sigma}$ is the level position on dot $i$, 
$\epsilon_{i,\sigma}=\epsilon_{i}$,  relative to
the chemical potential $\mu_i$, and $U_i$ is the intra-dot interaction.
\par

The coupling of the dots to the leads is described by a hybridization term,
\begin{eqnarray}
H_{{\rm c},i}
=\sum_{{\bf k},\alpha,i,{\sigma}}V_{{\bf k},\alpha,i}(c^{\dagger}_{{\bf k},i,\alpha,\sigma}d^{}_{i,\sigma} + \hc).
\end{eqnarray}
We will assume no energy dependence of the matrix elements but allow them to
differ in the different channels. We define the widths $\Gamma_{i}=\sum_\alpha\pi
V_{\alpha,i}^2\rho_c(0)$, where $\rho_c(0)$ is the conduction electron density of states at the Fermi level, as the constant energy scale for the hybridization. For transport close
to equilibrium only the  combination
$V_{i,s}c^\dagger_{{\bf k},i,s,\sigma}+V_{i,d}c^\dagger_{{\bf  k}, i,d,\sigma}$ 
couples to the dot states. We can therefore simplify the problem to two dots
and two itinerant channels.\par

Finally for  capacitively coupled  dots we assume 
 a repulsive interaction term between the charges on the
individual dots $U_{12}$,
\begin{equation}
  H_{12}= U_{12}\sum_{\sigma,\sigma'} n_{1,\sigma}n_{2,\sigma'}.
\end{equation}

If the dots are identical, with equal coupling to their baths
  and $U_{12}=U$,  then the model has  SU(4) symmetry.
This can be shown explicitly by combining the site and spin
indices, $(i,\sigma)\to\nu=((-1)^i+5/2+\sigma)$, where $\sigma=\pm 1/2$, so $\nu=1,2,3,4$, 
and express the Hamiltonian in terms of the creation and annihilation
operators $c^{\dagger}_\nu$ and $c^{}_\nu$. In the regimes with  integral
total occupation number $n_{\rm tot}=1,2,3$, which requires  strong
local interactions, this
SU(4) Anderson model can be mapped into an SU(4) Kondo model,
\begin{equation}
{\cal H}_{\rm K}= J \sum_{\nu,\nu',k,k'}Y_{\nu,\nu'}c^{\dagger}_{k',\nu'}c^{}_{k,\nu}
+ \sum_{\nu,k}\epsilon_{k}c^{\dagger}_{k,\nu}c^{}_{k,\nu},
\label{Kondo_model}\end{equation}
where the sum over $\nu=1,2, ... 4$, 
and for $U>D$, $J=
4|V|^2/U$ in the case with particle-hole symmetry. The operators $Y_{\nu,\nu'}$ obey
the SU(2n) commutation relations,
\begin{equation}
  [Y_{\nu,\nu'}, Y_{\nu'',\nu'''}]_-=Y_{\nu,\nu'''}\delta_{\nu',\nu''}-Y_{\nu'',\nu'}\delta_{\nu,\nu'''},
\end{equation}
with  $\sum_\nu Y_{\nu,\nu}=nI$, where $I$ is the identity operator. 
The case with  $n_{\rm tot}=1$ corresponds to the situation where
the occupation of the individual dots plays the role of a pseudospin,
and the $Y_{\nu,\nu'}$ operators correspond to the four dimensional 
(fundamental) representation of SU(4).  It is 
also a particular case of  the model introduced  by Coqblin and
Schrieffer \cite{CS69} to describe certain rare earth magnetic impurities (similarly for the case   $n_{\rm tot}=3$
in terms of holes). However, for the Kondo model with integral
occupation on each dot, such that $n_{\rm tot}=2$,
the operators  correspond to a six dimensional irreducible representation of SU(4) \cite{NCH10s}.\par
For two capacitively coupled dots  there is no symmetry or constraint such that
$U_{12}=U$, so we expect  the on-site
interaction $U$ to be significantly greater than the inter-site interaction  $U_{12}$. Estimates of the magnitude of the different interaction terms 
have been given in recent experimental work \cite{AKRCKSOG13}:
 $U_1\approx 1.2$meV, $U_2\approx 1.5$meV, $U_{12}\approx 0.1$meV,
 $\Gamma_1,\Gamma_2\approx 0.005-0.02$meV, so realistically there is no SU(4) symmetry in the 'bare' model.
We are concerned with the low energy regime, however,
where the effective or renormalized interactions determine
the behavior. There is the possibility that a new SU(4) symmetry  
can emerge on this scale, as originally predicted on the basis of scaling equations from the high energy regime  by Borda et al.\cite{BZHHD03}. In the next section we derive  precise criteria for a low energy
SU(4) fixed point in terms of renormalized parameters.\par

\section{Renormalized Parameters}

We start  from an exact expression for the  Fourier transform of the one-electron  Green's function $G_{i}(\omega)$
for
dot $i$,
\begin{equation}
G_i(\omega)=\frac{1}{\omega-\epsilon_i+i\Gamma_i-\Sigma_i(\omega)},
\end{equation}
where $\Sigma_i(\omega)$ is the proper self-energy. The corresponding
spectral density $\rho_i(\omega)$ is
\begin{eqnarray}
\rho_i(\omega)=-\frac{1}{\pi}\lim_{\delta\to 0}{\rm Im}\, G_i(\omega+i\delta)=\quad\quad\quad\quad\nonumber \\
\frac{1}{\pi}\frac{ \Gamma_i-\Sigma^I_i(\omega)}{(\omega-\epsilon_i-\Sigma^R_i(\omega))^2+( \Gamma_i-\Sigma^I_i(\omega))^2
},
\end{eqnarray}
where $\Sigma^R_i(\omega)$ and $\Sigma^I_i(\omega)$ are the real and imaginary parts of the self-energy.
We assume that the low energy behavior corresponds to a Fermi liquid so that  $\Sigma^I_i(\omega)$ is of
order $\omega^2$ as $\omega\to 0$. We can define a set of renormalized parameters \cite{Hew93,HOM04}, $\tilde\epsilon_i$, $\tilde\Gamma_i$,
$\tilde U_i$ and $\tilde U_{12}$, 
\begin{equation}\tilde\epsilon_i=z_i(\epsilon_i+\Sigma(0)),\quad \tilde\Gamma_i=z_i\Gamma_i,
\end{equation}
where $ z_i=1/(1-\Sigma'_i(0))$ is the wavefunction renormalization factor, and
\begin{equation}\tilde U_i=z^2_i\Gamma^{(4)}_{i,\uparrow, \downarrow}(0,0,0,0),\quad \tilde U_{12}=z_1z_2\Gamma^{(4)}_{12}(0,0,0,0),
\end{equation}
 where $\Gamma^{(4)}_{i,\uparrow, \downarrow}(\omega_1,\omega_2,\omega_3,\omega_4)$ and  $\Gamma^{(4)}_{12}(\omega_1,\omega_2,\omega_3,\omega_4)$ are the full four-vertices for on-site and inter-site scattering. 
If we replace the set of bare parameters of the  original Hamiltonian, $\epsilon_i,\Gamma_i, U_i,U_{12}$  with  the renormalized parameters, $\tilde\epsilon_i,\tilde\Gamma_i,\tilde U_i,\tilde U_{12}$, we obtain an effective Hamiltonian, which describes the
interacting quasiparticles which determine the low energy behavior \cite{NHCB13}. It should be noted, however, that the
quasiparticle interaction terms have to be normal ordered so that
these terms come into play only when more than one
quasiparticle is created, as the  ground state of the interacting system plays the role of a vacuum state. For calculations beyond the  Fermi liquid regime counter terms have to be explicitly included, and taken into account in a renormalized
perturbation expansion \cite{Hew93,Hew06}. 
The renormalized parameters that specify the quasiparticles and their interactions
  provide a complete guide to the low temperature behavior of the system. In particular, using these we can determine  precise criteria  for  an SU(4) symmetric low energy fixed point. We begin by noting a number of exact
relations which can be expressed in terms of these parameters. The well-known Friedel sum rule \cite{Lan66},
that gives the $T=0$ occupation number on each dot, can be expressed completely in terms 
of the parameters that specify the non-interacting quasiparticles,
\begin{equation}
n_i=1-\frac{2}{\pi}{\rm tan}^{-1}\left(\frac{\tilde\epsilon_i}{\tilde\Gamma_i}\right),
\label{fsr}
\end{equation}  
so the total occupation number of the two dots is given by  $n_{\rm tot}=n_1+n_2$.
Furthermore we have exact relations for several static response functions \cite{NHCB13}.
For example, the  total charge susceptibility $\chi_{c}$ of the double dot,
is given by
\begin{equation}
\chi_{c}=2\sum_{i=1,2}\tilde\rho^{(0)}_i(0)(1-\tilde U_i\tilde\rho^{(0)}_i(0)) -8\tilde U_{12}\tilde\rho^{(0)}_1(0)\tilde\rho^{(0)}_2(0),
\label{charge}
\end{equation}
and the total spin susceptibility $\chi_{s}$ by
\begin{equation}
 \chi_{s}=\frac{1}{2}\sum_{i=1,2}\tilde\rho^{(0)}_i(0)(1+\tilde U_i\tilde\rho^{(0)}_i(0)),
\end{equation}
where $\tilde\rho^{(0)}_i(\omega)$ is the free quasiparticle density of states for dot $i=1,2$,
\begin{equation}
\tilde\rho^{(0)}_i(\omega)=\frac{1}{\pi}\frac{\tilde\Gamma_i}{(\omega-\tilde\epsilon_i)^2+\tilde\Gamma_i^2}.
\end{equation}
The expression for the pseudospin susceptibility takes the form,
\begin{equation}
\chi_{ps}=\frac{1}{2}\sum_{i=1,2}\tilde\rho^{(0)}_i(0)(1-\tilde U_i\tilde\rho^{(0)}_i(0)) +2\tilde U_{12}\tilde\rho^{(0)}_1(0)\tilde\rho^{(0)}_2(0).
\end{equation}
From these we  can define Wilson ratios for the spin $R_s$  
 and pseudospin $R_{ps}$, as 
\begin{equation}
R_{s}=\frac {2\chi_s}{\tilde\rho^{(0)}_1(0)+\tilde\rho^{(0)}_2(0)}, \quad R_{ps}=\frac {2\chi_{ps}}{\tilde\rho^{(0)}_1(0)+\tilde\rho^{(0)}_2(0)}.
\label{sus}
\end{equation}
\par
\subsection{Conditions for an SU(4) Kondo fixed point}
\label{AA}

For identical dots, the condition,  $\tilde U_{12}=\tilde U_1=\tilde U_2$, is sufficient for the low energy fixed point of the double dot model to have SU(4) symmetry.  
 For non-identical dots we need to include explicitly the requirement, $\tilde\epsilon_1=\tilde\epsilon_2$ and $\tilde\Gamma_1=\tilde\Gamma_2$.  However, these extra conditions are not sufficient to ensure that  $\rho_1(\omega)=\rho_2(\omega)$ on the lowest energy scales. As $\rho_i(0)=z_i\tilde\rho_i(0)$, for non-identical dots 
 we have a further  condition $z_1=z_2$.\par  
 For an SU(4) {\em Kondo} fixed point
with  $n_{\rm tot}=1$ we need two extra conditions. From the Friedel sum rule (\ref{fsr}), for  $n_{\rm tot}=1$
we require $\tilde\Gamma=\tilde \epsilon$, or equivalently a phase shift $\eta=\pi/2-{\rm tan}^{-1}(\tilde\epsilon/\tilde\Gamma)=   \pi/4$.
For the Kondo regime we need to suppress the charge fluctuations and confine only 1 electron to the double dot. From Eqs. (\ref{charge}) and (\ref{sus}) this implies $R_s=R_{ps}=4/3$. With these conditions satisfied  there is universality
in terms of a single energy scale, the SU(4) Kondo temperature $T_{\rm K}^{(4)}$
which, for  $\tilde\rho^{(0)}_1(0)=\tilde\rho^{(0)}_2(0)=\tilde\rho^{(0)}(0)$, we can define by $T_{\rm K}^{(4)}=1/4\tilde\rho^{(0)}(0)$. \par\smallskip
These  conditions can be summarized as
\begin{eqnarray}&(i)&\quad \rho_1(\omega)=\rho_2(\omega)=\rho(\omega) \quad {\rm as}\quad\omega\to 0,\nonumber\\
&(ii)& \quad n_{\rm tot}=1,\nonumber\\
&(iii)&\quad R_s=R_{ps}=4/3.\nonumber
\end{eqnarray} 
If three of these conditions are satisfied we will describe the low energy fixed point as a universal Kondo SU(4)
fixed point with $n_{\rm tot}=1$. If condition (ii) and (iii) are satisfied, but (i) is only satisfied
at $\omega=0$, then we will describe the fixed point as a {\em restricted} SU(4) fixed point.
If only (i) and (ii) are satisfied, then there is no universal SU(4) fixed point; these two conditions can
be satisfied even for two isolated quantum dots, $U_{12}=0$. However, if the inter-dot interaction is 
large enough to  suppress significantly  the pseudospin fluctuations, say such that $4/3>R_{ps}\gtrsim 1 $,
we might describe the fixed point as an {\em approximate} SU(4) fixed point.

\par
\subsection{Calculation of renormalized parameters}
We can identify the low energy effective Hamiltonian, specified in terms of the renormalized parameters, as
the low energy fixed point of a numerical renormalization group (NRG) calculation
together with the leading order correction terms. This gives us an accurate way to deduce  the renormalized
parameters from the low energy many-body  excitations of an NRG calculation (for details see our earlier paper\cite{NHCB13}).
Hence, given  a set of `bare' parameters, which specify the full model Hamiltonian,  we can calculate the renormalized
parameters for the low energy effective model and test whether they are compatible with an emergent SU(4) fixed point.
\par

There are certain obvious conditions that have to be fulfilled to achieve an SU(4) Kondo state for
this double dot system. A single electron has to be localized on the two dots. The on-site interaction $U_i$ on a single dot must be large compared with
the bath coupling $\Gamma_i$  to suppress fluctuations
giving double occupancy. This only restricts the occupation of a dot to the range $0\le n_i \le 1$,
so inter-site interaction $U_{12}$ has to be large enough to suppress double occupancy of the combined system.
Ideally the two quantum dots should also be identical, which can be difficult to achieve experimentally.
The energy level $\epsilon_i$ on each dot can be controlled by a gate voltage on each dot  not only  to adjust
the electron occupation on the dot but also to match the two dots. 
 Any difference in the on-site interaction $U_i$
between the dots is unlikely to be important as long as they are both large enough to suppress any significant
double occupation. As pointed out by Tosi at al. \cite{TRA15} it can be difficult to match the couplings between
the baths and dots such  $\Gamma_1=\Gamma_2$.
 The value of $\Gamma$ is a very significant one in determining the
degree of renormalization, and the Kondo temperature for a single dot depends exponentially on this quantity. 
They argue, however, that when one takes account of Haldane scaling \cite{Hal78}, which   gives an  effective shift
of the bare levels on each dot  $\epsilon_i\to \epsilon^*_i$, so the effect of  the 
 difference in $\Gamma_1$ and $\Gamma_2$ is translated into a difference in the effective levels on the dots.
This difference can then  be eliminated by adjusting the gate voltages on each dot so that symmetry is effectively restored.
 Their suggestion was supported
by explicit calculations using the non-crossing approximation (NCA).
A drawback of the NCA method, however, is that it is difficult
to apply to the model with finite values of $U$ and $U_{12}$,
so their calculations were limited to the case  with $U\to \infty$ and $U_{12}\to\infty$.
 A further limitation is that the NCA  breaks down on scales much less than the Kondo temperature, so that it cannot describe completely the Fermi liquid regime. \par

Our earlier calculations \cite{NHCB13} were for a double dot  model with identical
dots, constrained such that $n_{\rm tot}=1$.
We addressed the question:  How large do the on-site and inter-site interactions
have to be to achieve a universal SU(4) Kondo state? 
We found it was difficult to achieve such a state  with the physically appropriate
limitation $U_{12}<U$,
even if  both $U/\pi\Gamma>3$ and $U_{12}/\pi\Gamma>3$.
 Only in the 
very limited situation with $U> U_{12}>D$, could 
the requirement $\tilde U=\tilde U_{12}$ be asymptotically satisfied. Here we re-examine the
question as to how close we can approach an SU(4) point for a range of strengths
of the interaction parameters $U_i$ and $U_{12}$. We will also compare
the characteristic features of an SU(4) point for the double dot with
$n_{\rm tot}=1$, due to spin and pseudospin, with that for $n_{\rm tot}=2$
due to spin alone. However, we begin by examining a model with 
different values of $\Gamma_i$ to see whether the symmetry restoration
mechanism of Tosi et al. still holds for finite values of the interaction
parameters $U_i$ and $U_{12}$.
\section{Unequal couplings: Symmetry restoration?}

 We start first of all with a choice of parameters for the two dots such that  $U_1=U_2=U_{12}$, so that they differ only their          couplings to the bath  $\Gamma_1\ne \Gamma_2$ and in
 their  one-electron levels $\epsilon_1\ne\epsilon_2$.  We assume that we can independently adjust the  two gate voltages   to ensure both a total occupation $n_{\rm tot}=1$ and compensate for  the difference in the couplings. As noted earlier in section \ref{AA}, the conditions,
$\tilde\epsilon_1=\tilde\epsilon_2$, $\tilde\Gamma_1=\tilde\Gamma_2$ and  $\tilde U_1=\tilde U_2=\tilde U_{12}$,
are not sufficient in general to ensure $\rho_1(\omega)=\rho_2(\omega)$, so there is the additional requirement, $z_1=z_2$.  However, from the definition $\tilde\Gamma_i=z_i\Gamma_i$, the conditions, $\tilde\Gamma_1=\tilde\Gamma_2$  and  $z_1=z_2$, are only compatible if $\Gamma_1=\Gamma_2$.
We conclude that, if  $\Gamma_1\ne\Gamma_2$, we cannot satisfy all the conditions for strict low energy SU(4) symmetry. \par

 However, if we relax these conditions and require  SU(4) symmetry only at $\omega=0$, corresponding to what we have described as a restricted SU(4) fixed point. This would require $\tilde\rho^{(0)}_1(0)=\tilde\rho^{(0)}_2(0)$ and
$z_1=z_2$ so that  $\rho_1(0)=\rho_2(0)$.  We might be able to satisfy these conditions 
 in a model with  $\Gamma_1\ne\Gamma_2$. We now put these ideas to the test with some particular examples, using the NRG
to calculate the renormalized parameters.
\par 
 
\begin{table}[h]
\caption{The interaction parameters are in units of the half-bandwidth $D=1$, and  the Wilson
ratios for spin $R_s$ and pseudo-spin $R_{ps}$  are evaluated at point where the difference in dots levels $\delta_c$ gives a maximum  $R_{ps}$ (local minimum in $R_s$).}
\centering
\begin{tabular}{|c|c|c|c|c|c|c|c|}
\hline
 $U_1$  & $U_2$   & $U_{12}$ &    $\delta_c$        & $z_2/z_1$ & $R_s$ & $R_{ps}$ & $T_{\rm K}$ \\
 \hline
$0.5$  & $0.5$  &  $0.5$  &  $2.96\times 10^{-4}$  & $1.05$  & $1.329$ & $1.340$ & $1.0\times 10^{-10}$\\
 \hline
 $0.12$ & $0.12$ & $0.12$  &  $2.05\times 10^{-4}$ & $1.03$  & $1.330$ & $1.340$ & $2.5\times 10^{-5}$\\
 \hline
 $0.05$ & $0.05$ & $0.05$  & $-2.2\times 10^{-4}$ & $1.02$  & $1.329$ & $1.336$ & $6.3\times 10^{-4} $\\  
\hline
 $ 5.0$  & $5.0$  & $3.0$  &  $6.11\times 10^{-4}$ & $1.05$  & $1.332$  &  $1.336$ & $2.4\times 10^{-8}$\\ 
 \hline
 $0.05$ & $0.05$ & $0.03$ & $-2.14\times 10^{-5}$ & $ 1.05$ & $1.475$  & $1.017$   & $1.3\times 10^{-3}$\\
 \hline
 $1.0$  & $0.8$  & $0.1$  & $-4.32\times 10^{-4}$ &  $1.13$   & $1.439$ & $1.121$ & $3.5\times 10^{-5}$\\
 \hline
 $0.5$  & $0.4$  & $0.04$ & $-6.66\times 10^{-4}$ &  $1.12$   & $1.551$ & $0.889$ & $6.1\times 10^{-4}$\\ 
\hline

\end{tabular}

\label{tab}
\end{table}
We start with fixed values of $U_1$, $U_2$, $\Gamma_1$, $\Gamma_2$, and
 values of  $\epsilon_1$ and  $\epsilon_2$
such that we are in a localized regime with $n_{\rm tot}\sim 1$.
In all cases, unless mentioned otherwise, we take $\pi\Gamma_1=0.01$ and $\pi\Gamma_2=0.007896$
 (the energy scale is set by the half-bandwidth $D=1$).
 We then vary the value  $\delta\epsilon_{12}=\epsilon_1-\epsilon_2$, maintaining $n_{\rm tot}\sim 1$, and calculate the set of
renormalized parameters, $\tilde\Gamma_i$, $\tilde\epsilon_i$, $\tilde U_i$ and $\tilde U_{12}$,
as a function of $\delta\epsilon_{12}$. From these results we can deduce $\tilde\rho^{(0)}_i(0)$, $z_i$, and the Wilson ratios
for the spin and isospin, $R_s$ and $R_{ps}$. We define $\delta_c$ as the value of $\delta\epsilon_{12}$
corresponding the maximum value of  $R_{ps}$, which is the point corresponding to the best
approximation to an SU(4) fixed point. It will be convenient to use the
 variable $\delta\epsilon$ defined by
 \begin{equation}\delta\epsilon=\delta\epsilon_{12}-\delta_c,
\end{equation} as a measure of the energy difference away from this point.
 The results for  $\delta_c$  for several parameters sets 
are given in Table I together with the values at this point for $z_2/z_1$,  $R_s$ and $R_{ps}$. The quantities give us some measure of the  proximity to a precise fixed point at $\omega=0$,
which would correspond to $z_2/z_1=1$,  $R_s=R_{ps}=4/3$. Before commenting on the general trends, we look at some of the results in detail.
\par
  
\begin{figure}[!htbp]
\begin{center}

{\includegraphics[width=0.75\linewidth]{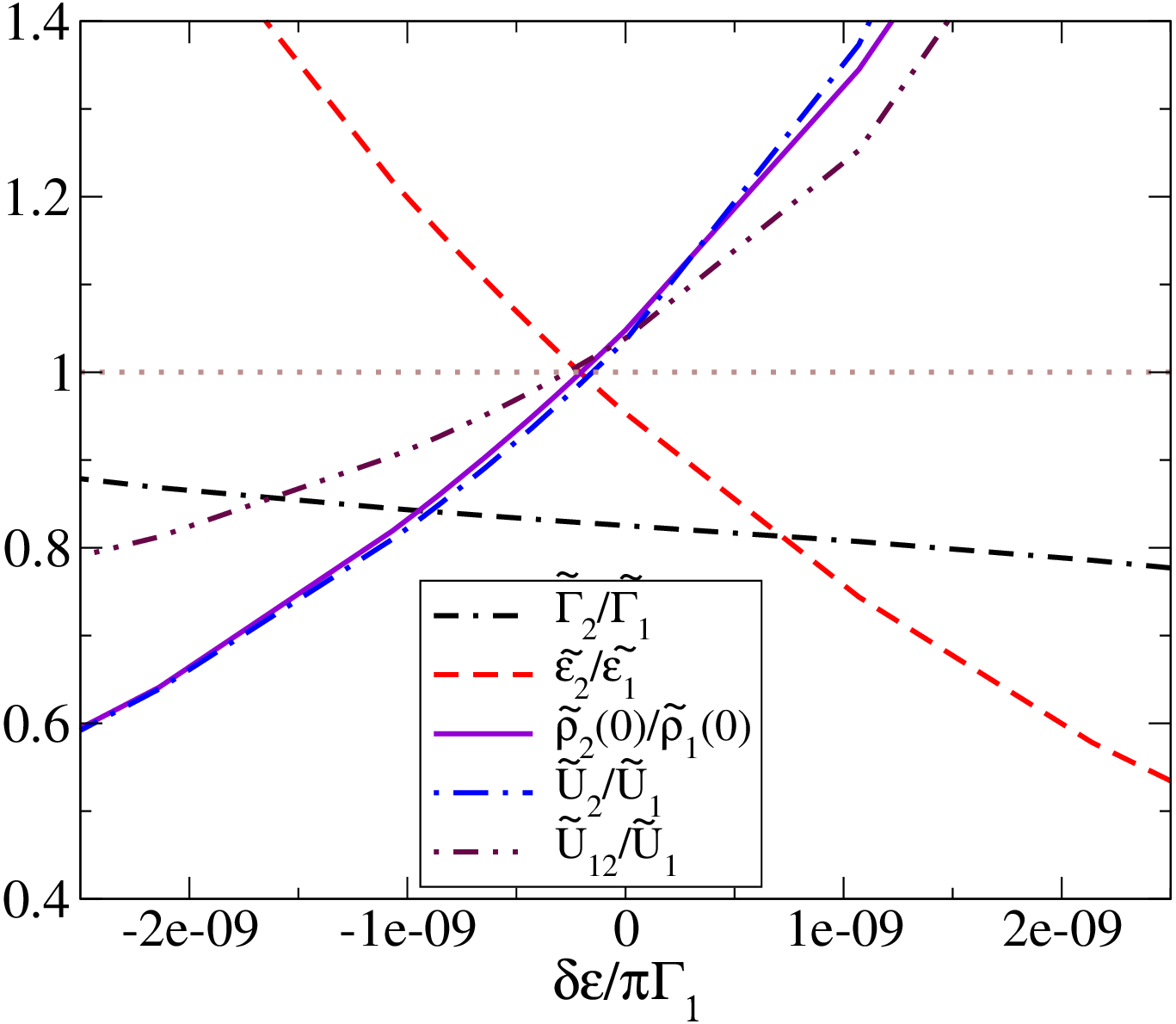}}
\caption
{ The ratios of renormalized parameters, $\tilde\Gamma_2/\tilde\Gamma_1$,
 $\tilde\epsilon_2/\tilde\epsilon_1$, $\tilde\rho^{(0)}_2(0)/\tilde\rho^{(0)}_1(0)$, $\tilde U_2/\tilde U_1$
and  $\tilde U_{12}/\tilde U_1$ as a function of $\delta\epsilon/\pi\Gamma_1$ for
 $U_1=U_2=U_{12}=0.5$, $\epsilon_1=-0.093$.}
\label{nren2}
\vspace{0.5cm}
{\includegraphics[width=0.73\linewidth]{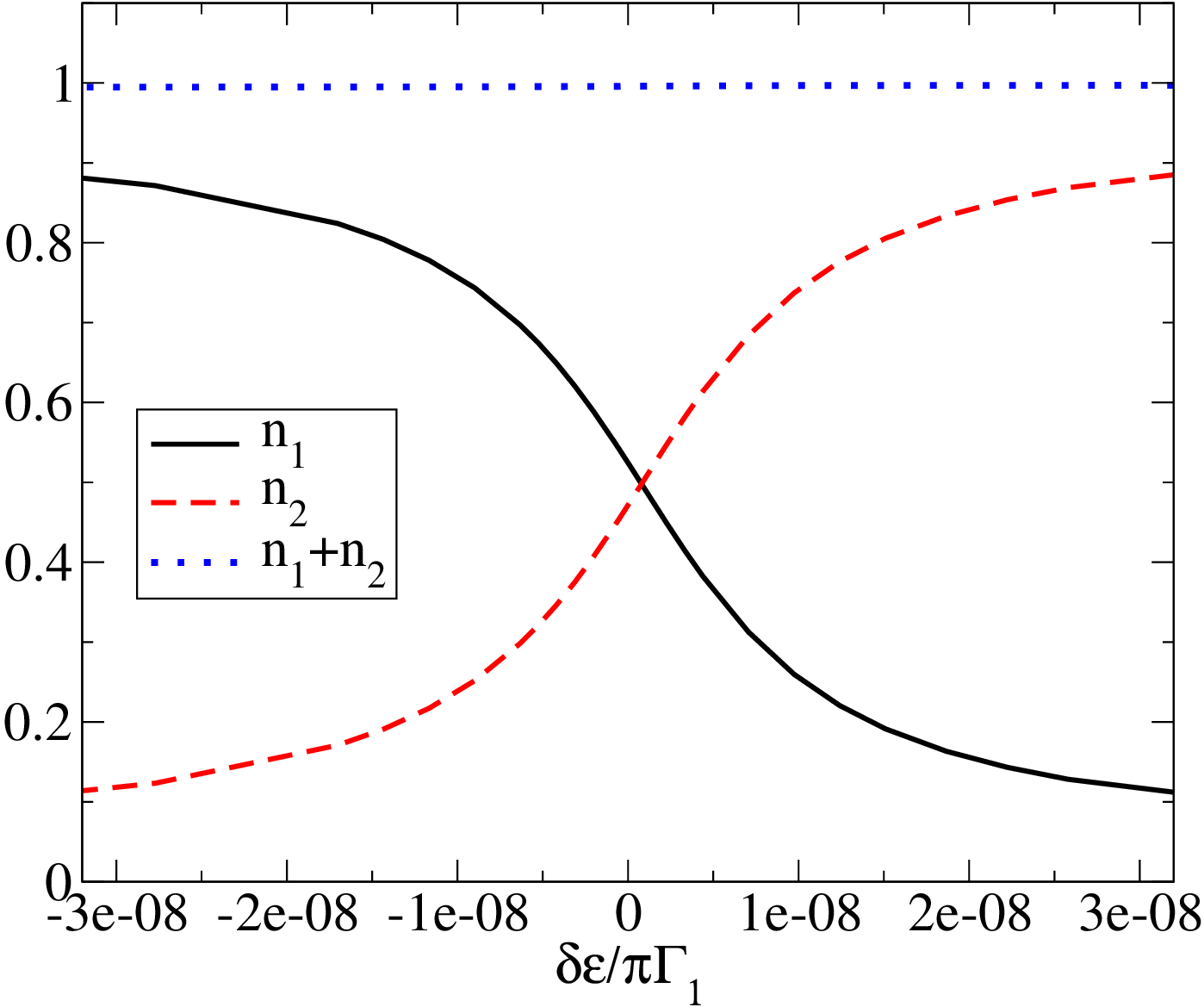}}
 \caption
{(Color online) The occupation numbers  on the individual dots, $n_1$,
$n_2$, and their sum, 
as a function of $\delta\epsilon/\pi\Gamma_1$ for the parameter set in Fig. \ref{nren2}.}
\label{ntot2}

\end{center}
\end{figure}



\noindent
\begin{figure}[!htbp]
\begin{center}
{\includegraphics[width=0.75\linewidth]{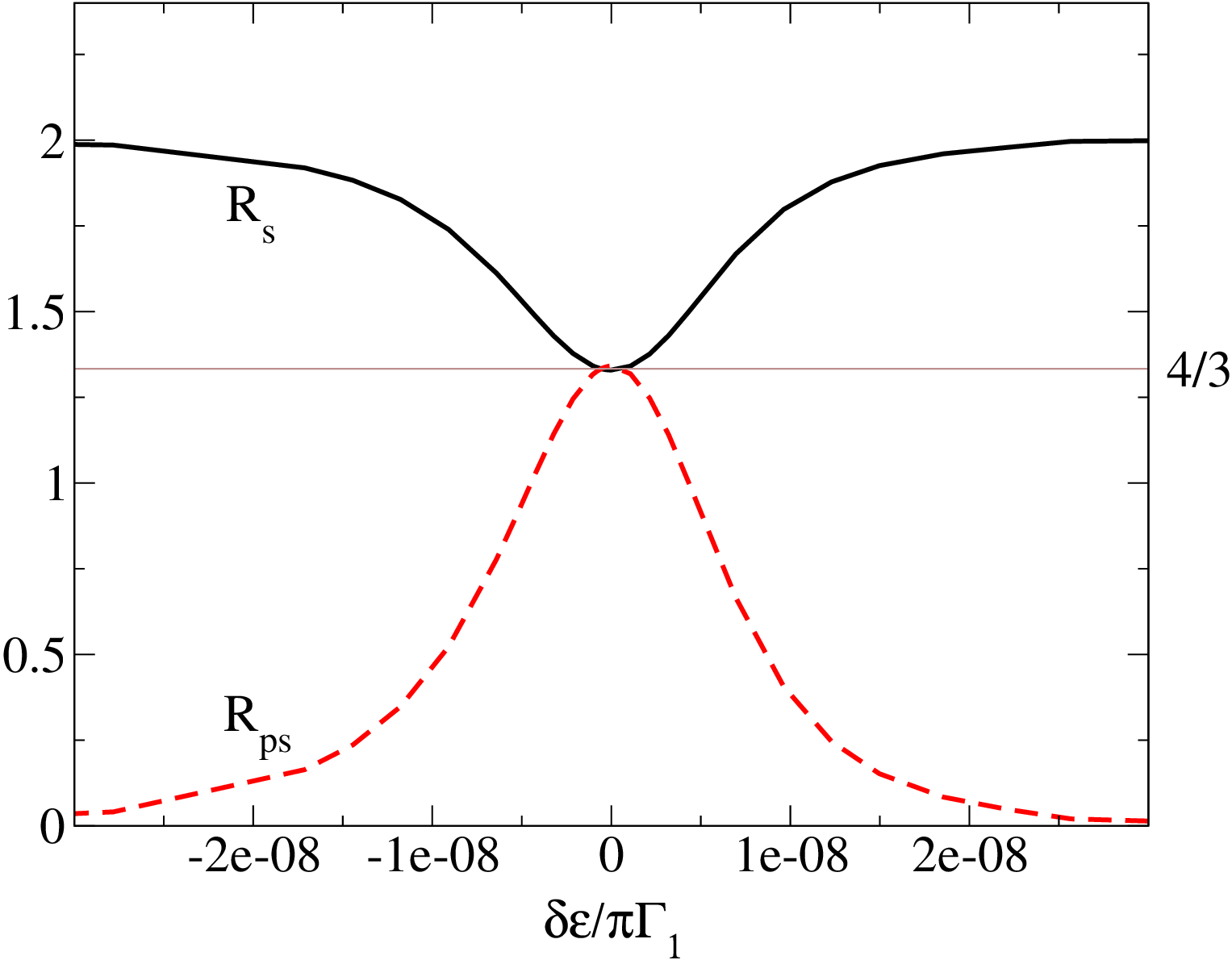}}
\caption{(Color online)  The Wilson ratios for the spin and isospin, $R_s$ and $R_{ps}$,
as a function of $\delta\epsilon/\pi\Gamma_1$  for the parameter set in Fig. \ref{nren2}.}
\label{nwr2}
\vspace{0.5cm}
{\includegraphics[width=0.6\linewidth]{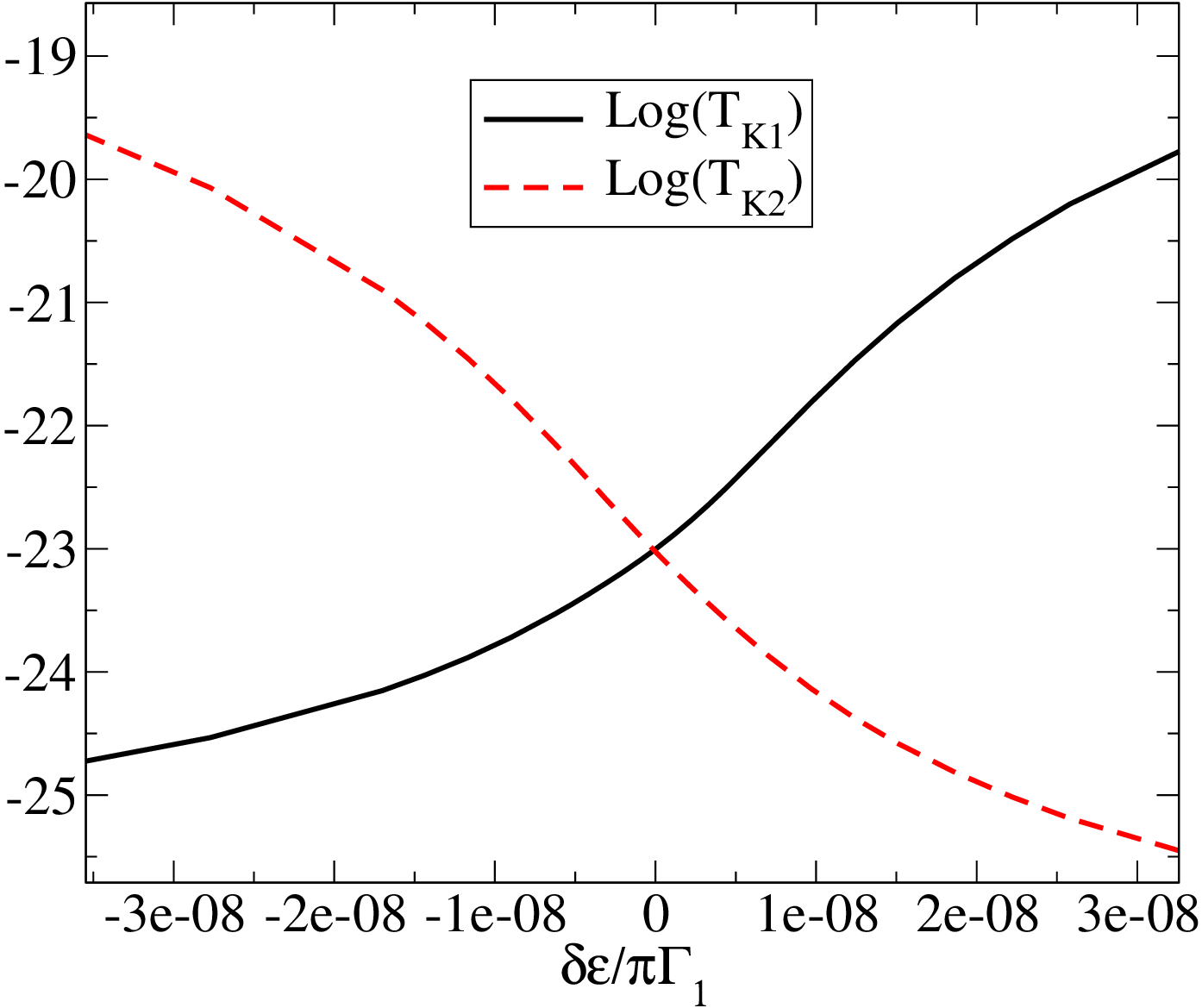}}
\caption
{(Color online)The logarithms of the Kondo temperatures for the individual dots, $T_{\rm K,1}$ and  $T_{\rm K,2}$,  as a function of $\delta\epsilon/\pi\Gamma_1$ 
for the parameter set in Fig. \ref{nren2}.}
\label{nstar2}
\end{center}
\end{figure}

\noindent

\begin{figure}[!htbp]
\begin{center}
{\includegraphics[width=0.72\linewidth]{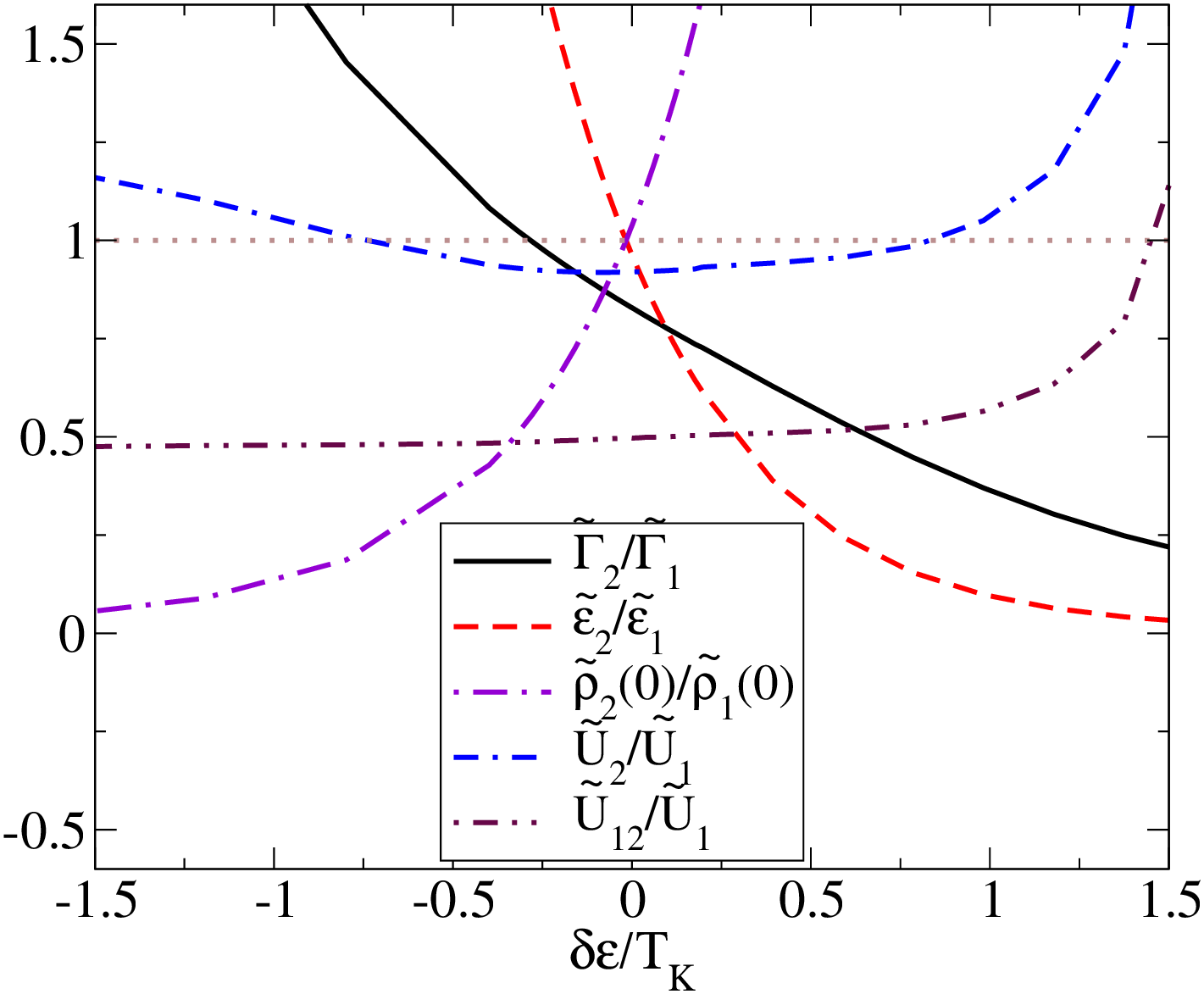}}
\caption{ The ratios of renormalized parameters, $\tilde\Gamma_2/\tilde\Gamma_1$,
 $\tilde\epsilon_2/\tilde\epsilon_1$, $\tilde\rho^{(0)}_2(0)/\tilde\rho^{(0)}_1(0)$, $\tilde U_2/\tilde U_1$
and  $\tilde U_{12}/\tilde U_1$as a function of  $\delta\epsilon/T_{\rm K}$ for the parameter set $U_1=U_2=0.05$, $U_{12}=0.03$ and $(\epsilon_1+\epsilon_2)/2=-0.01468$. }
\label{nren5}
\vspace{0.5cm}
{\includegraphics[width=0.7\linewidth]{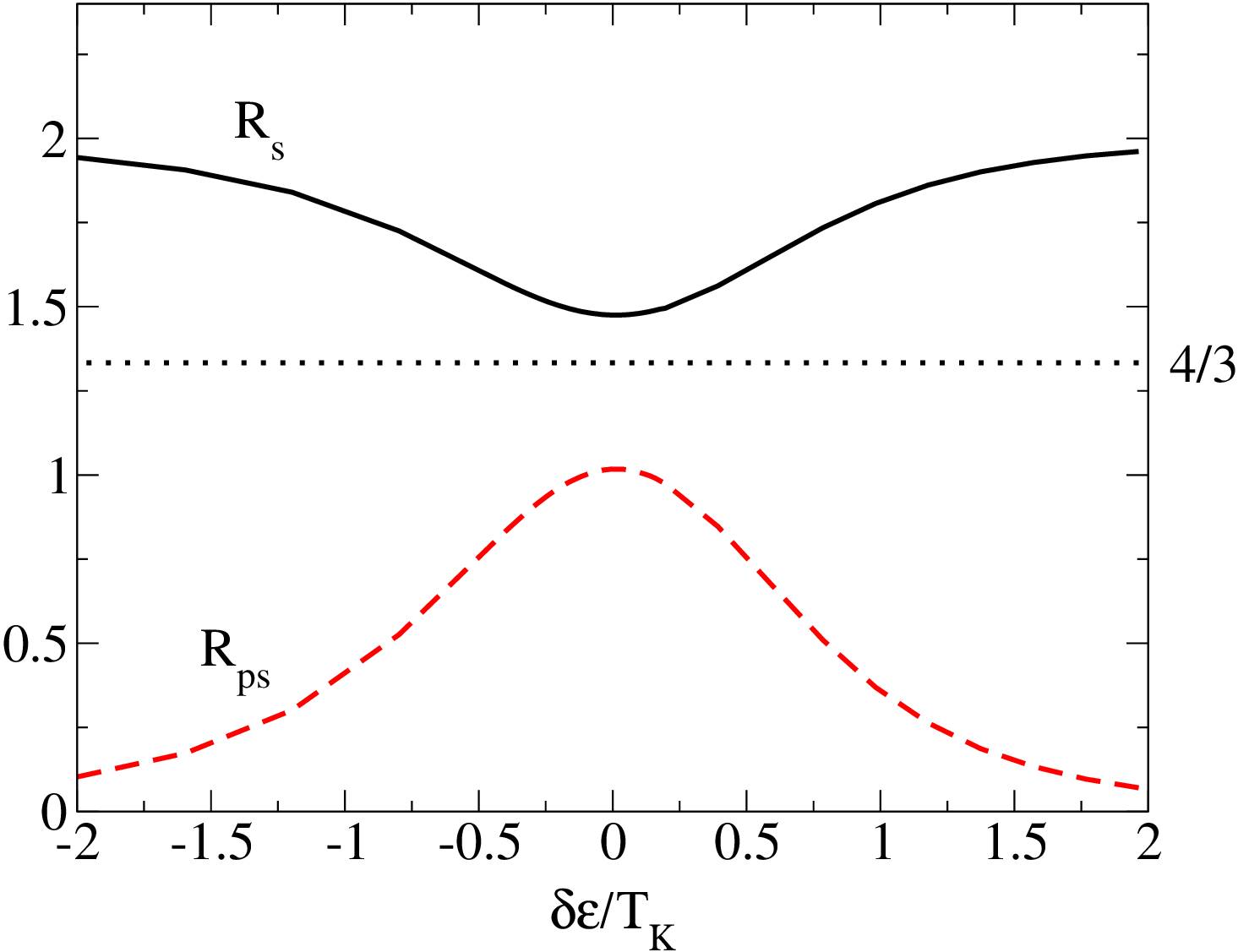}}
\caption
{(Color online) The Wilson ratios for the spin and isospin, $R_s$ and $R_{ps}$,
plotted as a function of  $\delta\epsilon/T_{\rm K}$ for the parameter set in Fig. \ref{nren5}
}
\label{nwr5}
\end{center}
\end{figure}


\noindent

We first of all consider the case with  $U_1=U_2=U_{12}=0.5$, where we take the value $\epsilon_1=-0.093$ 
and adjust $\epsilon_2$. We cover a parameter range in which a  single electron is confined to the two dots. This should be a favourable case to find a point with approximate  SU(4) symmetry as we have taken  
the interdot repulsion $U_{12}=U_1=U_2$
and with a value comparable with $D$, similar to the situation
considered by Tosi et al.  The condition  $U_{12}= U_1=U_2$, however,
 does not ensure that  $\tilde U_{12}=\tilde U_1=\tilde U_2$  because the degree of renormalization of the spin and pseudospin fluctuations, as we shall see, can differ in general.
We present results for the parameter ratios, $\tilde\Gamma_2/\tilde\Gamma_1$, $\tilde\epsilon_2/\tilde\epsilon_1$, $\tilde\rho^{(0)}_2/\tilde\rho^{(0)}_1$,  $\tilde U_2/\tilde U_1$ and  $\tilde U_{12}/\tilde U_1$, which are plotted in Fig.\ref{nren2}   as a function of $\delta\epsilon/\pi\Gamma_1$ ($\delta\epsilon$ is measured relative to $\delta_c$, which takes the value
$\delta_{c}=2.960151362\times 10^{-4}$). For complete
SU(4) symmetry all these curves should intersect at the same point with a value of 1. There is a clustering of intersections about this point so to a good approximation this is the case, the exception being the ratio $\tilde\Gamma_2/\tilde\Gamma_1$.
 However, we have argued that for SU(4) symmetry at $\omega=0$, it is not necessary for this ratio
to be equal to 1, but we do require $z_2/z_1\sim 1$. The ratio  $\tilde\Gamma_2/\tilde\Gamma_1\sim 0.83$ in this regime and from  $z_2/z_1=(\tilde\Gamma_2/\tilde\Gamma_1)(\Gamma_1/\Gamma_2)$
we deduce  $z_2/z_1\approx 1.05$, which is close enough for a resticted SU(4) point. In Fig. \ref{ntot2}  the values of occupation numbers on the dots $n_1$ and $n_2$, together with their sum
$n_{\rm tot}$ are plotted over the same range, verifying that we are covering a range with $n_{\rm tot}$ very close
 to the value 1. Away from the restricted SU(4) fixed point   we see that the ratio  $\tilde U_{12}/\tilde U_1$
differs from the ratio $\tilde U_{2}/\tilde U_1$ even though we have taken, $U_{12}= U_1$, reflecting the fact the on-site and inter-site renormalizations
differ in general.
\par 

\begin{figure}[!htbp]
\begin{center}
{\includegraphics[width=0.75\linewidth]{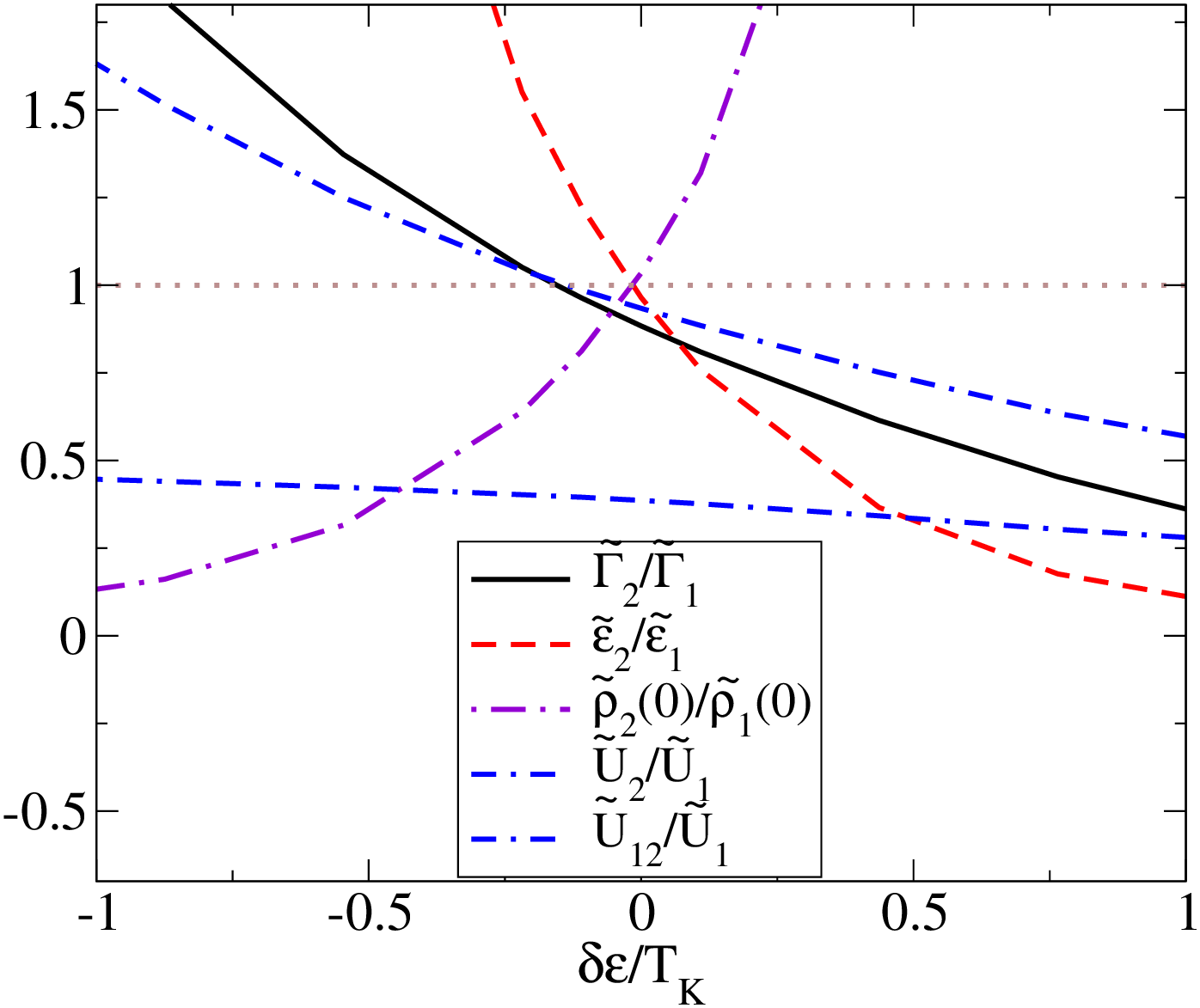}}
\caption
{(Color online) The ratios of renormalized parameters, $\tilde\Gamma_2/\tilde\Gamma_1$,
 $\tilde\epsilon_2/\tilde\epsilon_1$, $\tilde\rho^{(0)}_2(0)/\tilde\rho^{(0)}_1(0)$, $\tilde U_2/\tilde U_1$
and  $\tilde U_{12}/\tilde U_1$ for the parameter set $U_1=0.5$, $U_2=0.4$, $U_{12}=0.04$ and $(\epsilon_1+\epsilon_2)/2=-0.0205$.}
\label{nren6}
\vspace{0.5cm}
{\includegraphics[width=0.7\linewidth]{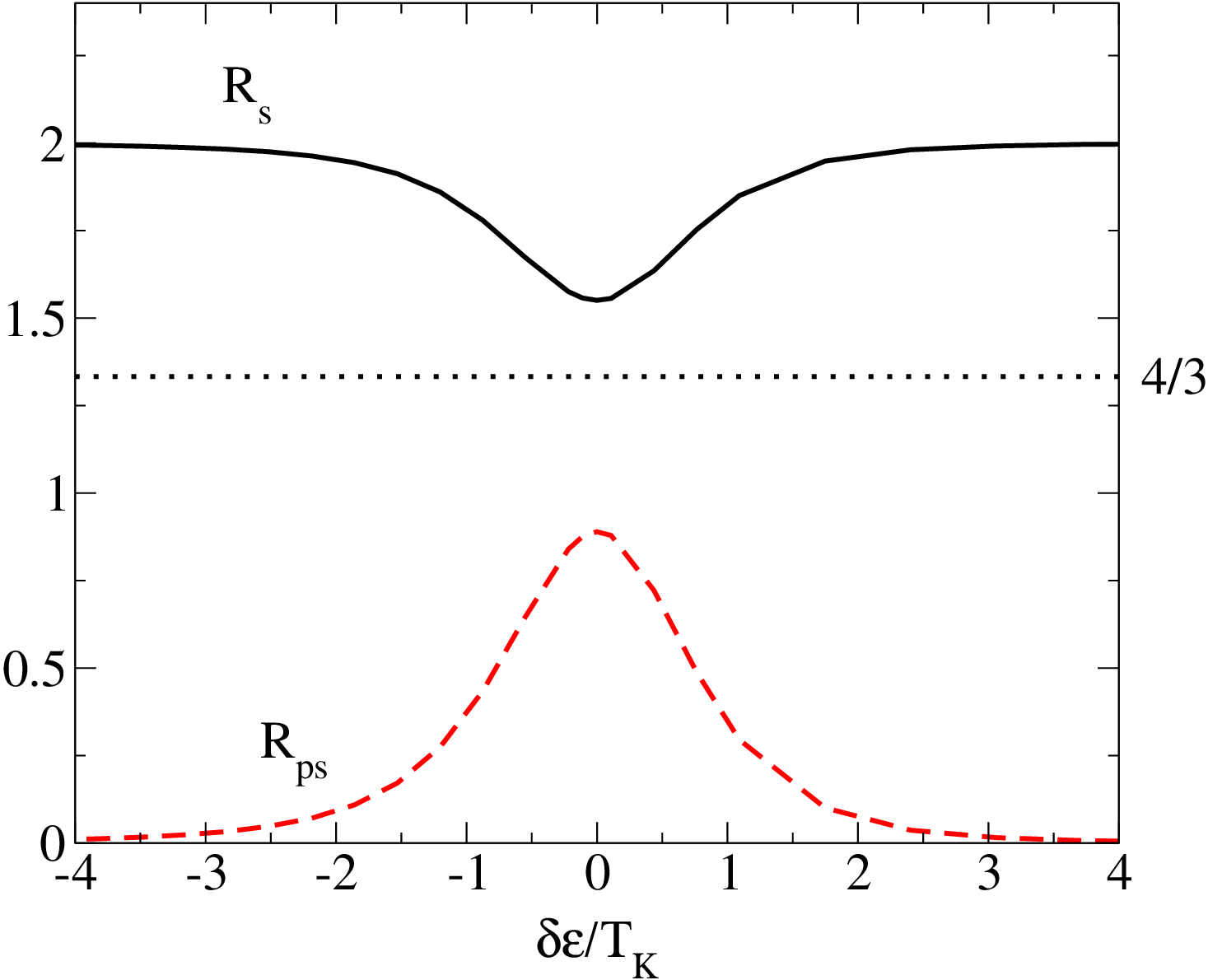}}
\caption
{(Color online) The Wilson ratios for the spin and isospin, $R_s$ and $R_{ps}$,
plotted as a function of  $\delta\epsilon/T_{\rm K}$ for the parameter set in Fig. \ref{nren6}.}
\label{nwr6}
\end{center}
\end{figure}

\noindent

 The corresponding values of the Wilson
ratios for the spin $R_s$ and isospin $R_{ps}$  are shown in Fig. \ref{nwr2} 
plotted against $\delta\epsilon/\pi\Gamma_1$.
At $\delta\epsilon=0$ both these ratios are almost equal   $4/3$ as expected at an  SU(4) fixed point. Away from the SU(4) region it can be seen  that the Wilson ratio for the spin $R_s$ takes a value 2, which corresponds to a regime
in which a single electron is confined to just one of the dots. As a result in this regime there are few inter-dot
 fluctuations so the pseudospin Wilson ratio $R_{ps}$ drops to almost zero.\par
We can define  Kondo temperatures  for the individual dots   $T_{{\rm K}i}$ via $T_{{\rm K}i}=1/4\tilde\rho^{(0)}_i(0)$ and  an approximate SU(4) Kondo temperature $T^{(4)}_{{\rm K}}$ as the point where these two
 Kondo temperatures are equal,  $T_{{\rm K}1}=T_{{\rm K}2}=T^{(4)}_{{\rm K}}$.
Even though this gives a single value it does not imply a Kondo regime and  universality unless all the
renormalized parameters can be expressed in terms of this single energy scale. 
 Away from the  point where  $T_{{\rm K}1}=T_{{\rm K}2}$ these two temperatures differ widely as can be seen in Fig. \ref{nstar2} , where we plot both ${\rm log}(T_{{\rm K}1})$
and ${\rm log}(T_{{\rm K}2})$ as a function of $\delta\epsilon/\pi\Gamma_1$. The value of $T_{\rm K}$
at the SU(4) fixed point is very small  $T_{\rm K}= 1.01\times 10^{-10} $, due to the large  values taken
for the interactions relative to the hybridization widths.\par
We give  some of the results for two more parameter sets with $U_1=U_2=U_{12}$ for $U_1/\pi\Gamma_1=12$
and $U_1/\pi\Gamma_1=5$ in Table I. These interaction terms are much reduced from the set we have just considered in detail, but they are  still in the Kondo regime with the inter-dot charge fluctuations
suppressed. In both cases there is a point corresponding to an restricted  SU(4) fixed point but, with
the reduced values of the interaction parameters, the Kondo temperatures are significantly bigger.\par
In the next parameter set given in Table I, all the interactions terms are significantly
larger than the bandwidth, $U_1/D=U_2/D=5$, $U_{12}/D=3$ ($D=1$), but the inter-site term is reduced
relative to the on-site interactions, reflecting a more realistic double dot situation. We see  again that there is a good restricted SU(4) fixed point. So the reduction in the value of $U_{12}$ relative to $U_1$ and $U_2$ does not at first sight have made any significant difference. However, the value of
the Kondo temperature is very much bigger than for the set $U_1=U_2=U_{12}=0.5$ with smaller values
of the interactions. \par
In the next set in Table I we have the results for a case where the interaction parameters are reduced to be much less than the bandwidth,  $U_1/D=U_2/D=0.05$, $U_{12}/D=0.03$ ($(\epsilon_1+\epsilon_2)/2=-0.01468$) but still in the strong correlation regime. The results for the ratios of the renormalized parameters and  the Wilson ratios are shown  in Figs. \ref{nren5} and \ref{nwr5} plotted  as a function of $\delta\epsilon/T_{\rm K}$,
with $T_{\rm K}$ as defined earlier as the value where $T_{{\rm K}1}= T_{{\rm K}2}=T_{\rm K}$. We see that these results are in marked contrast to the similar set with $U_1=U_2=U_{12}=0.05$,
which differs from this set only in the value of $U_{12}$.
A comparison of the results in Fig. \ref{nren5}  with those Fig. \ref{nren2}  shows that, though there is an approximate point where most of the ratios take a value of order 1, the ratio $\tilde U_{12}/\tilde U_1$ falls well below this point. As a consequence we see in  Fig. \ref{nwr5} that though there
is a peak in $R_{ps}$ and a dip in $R_s$ neither of these attain  the required value of $4/3$ for a universal
SU(4) Kondo
fixed point.  As the peak in  $R_{ps}$ is greater than 1, we can classify the fixed point as an approximate
SU(4) Kondo fixed point.
\par
We see similar results in Figs. \ref{nren6}  and \ref{nwr6} corresponding to the parameter set,
$U_1=0.5$, $U_2=0.4$, $U_{12}=0.04$ ($(\epsilon_1+\epsilon_2)/2=0.007896$). In this case the ratio $z_2/z_1=1.12$ giving a significant deviation
from 1. Also as $R_{ps}<1$, by our criteria the fixed point in this case does not qualify as an 
approximate SU(4) fixed point. 
The results are very similar in the other case considered; $U_1=1.0$, $U_2=0.8$, $U_{12}=0.1$, but with the larger value of $U_{12}$  we find an approximate SU(4) point as $R_{ps}>1$.
\par

We conclude from these examples, with $U_1=U_2=U_{12}$ but with  asymmetry of the coupling so
that $\Gamma_1\ne\Gamma_2$, that it is  possible in the Kondo regime with $n_1+n_2\sim 1$, to achieve
to a good approximation a restricted  energy SU(4) fixed point by adjusting the difference in energy levels,
$\epsilon_1-\epsilon_2$, in line with the conclusions of Tosi et al. The shifts $\delta_c$ required
to obtain this point in all our examples is of the same order of magnitude $\sim 10^{-4}$. Tosi et al. were able to relate this shift quantitatively to the formula for the Haldane scaling.  This we were not able to do here, but for  $U\to \infty$ and $U_{12}\to \infty$ there is  only one relevant cut-off $C=|\epsilon|$. With finite and different  values of both $U$ and $U_{12}$ in our examples, the charge scaling regimes will have different lower cut-offs so no simple universal formula is likely to apply.\par

 We also found a restricted SU(4) fixed point 
for $U_{12}<U_1$, $U_{12}<U_2$ but only for  $U_{12}$  greater than the conduction band width $D$,  
ie.  $U_{12}>1$. However, for  $U_{12}/\pi\Gamma_1\gg 1$ but $U_{12}<1$,
we find only a non-universal approximate SU(4) fixed point. This indicates at the low energy regime the inter-dot and on-site
interactions act differently.
\par

\subsection{NRG Spectral Densities}

\begin{figure}[!htbp]
\begin{center}
{\includegraphics[width=0.8\linewidth]{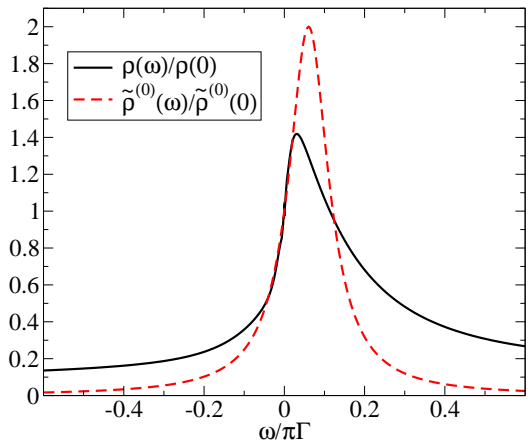}}
\caption
{(Color online). Plots of $\rho(\omega)/\rho(0)$ and  $\tilde\rho^{(0)}(\omega)/\tilde\rho^{(0)}(0)$
for an SU(4) model with $n_{\rm tot}=1$ for $U_{12}=U_1=U_2=0.05$, $\pi\Gamma_1=\pi\Gamma_2=0.01$.}
\label{set3}
\end{center}
\end{figure}

We now examine some of the results at a restricted
SU(4) fixed point with $n_{\rm tot}\sim 1$ on higher energy scales. We look
first of all at a case with identical dots and hybridizations, $\pi\Gamma_1=\pi\Gamma_2=0.01$,
and $U_1=U_2=U_{12}=0.12$, which has SU(4) symmetry on all energy scales.
  We expect the free quasiparticle expression $z\rho^{(0)}(\omega)$ to give a good approximation to the full spectral density $\rho(\omega)$ in the low frequency regime near the Fermi level. To test this we plot the ratios $\rho(\omega)/\rho(0)$ and  $\tilde\rho^{(0)}(\omega)/\tilde\rho^{(0)}(0)$ as a function of $\omega$ in Fig. \ref{set3}. We see that this is indeed the case and  the quasiparticle result accurately 
reproduces the sharp rise in the spectral density  in the low energy regime.
\par  
\bigskip
\begin{figure}
\begin{center}
{\includegraphics[width=0.8\linewidth]{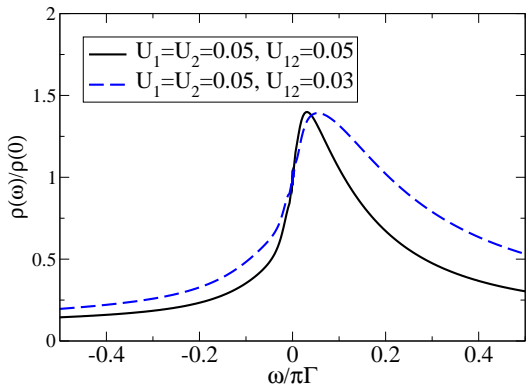}}
\caption
{(Color online). 
 Plots of  $\rho(\omega)/\rho(0)$ for the parameter set as in \ref{set3}, and for almost the same set except with
a reduced value of $U_{12}=0.03$. 
} 
\label{set34}
\end{center}
\end{figure}
\noindent
\begin{figure}[!htbp]
\begin{center}
\includegraphics[width=0.75\linewidth]{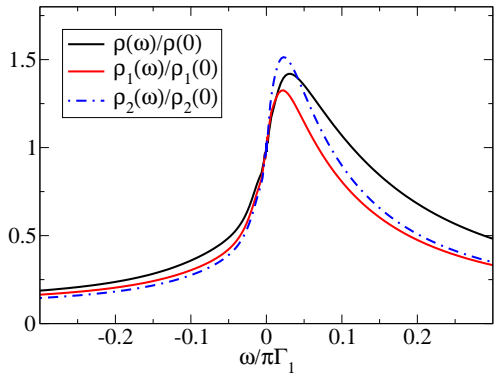}
\caption
{(Color online)  A plot of $\rho(\omega)/\rho(0)$ for the parameter set given in Fig. \ref{set3} 
and $\rho_1(\omega)/\rho_1(0)$ and $\rho_2(\omega)/\rho_2(0)$ for a model with the same interaction parameters ($U_{12}=U_1=U_2=0.05$) but with  $\pi\Gamma_1=0.01$ and $\pi\Gamma_2=0.007896$. }
\label{set36}
\end{center}
\end{figure}
\noindent

In Fig. \ref{set34}  we give two plots of $\rho(\omega)/\rho(0)$  for identical dots $n_1=n_2=1/2$, one for the set $U_1=U_2=U_{12}=0.05$ and the other for  $U_1=U_2=0.05$, $U_{12}=0.03$. The first corresponds to an
SU(4) fixed point with a Wilson ratio, $R_s=R_{ps}=1.329$, the second set with  $R_s=1.48$ and  $R_{ps}=1.005$,
so  corresponds only to an approximate SU(4) fixed point. Though the spectral densities have the same value at the Fermi level,
$\rho(0)\sim 1/2\pi\Gamma$, they deviate away from this point. Such a deviation would be expected even
if  both sets corresponded to an SU(4) fixed point because they would have different values of $T_{\rm K }^{(4)}$, but the comparison does reveal that the reduction in $U_{12}$ 
 significantly affects the spectrum on all energy scales.\par
In Fig. \ref{set36} we compare $\rho(\omega)/\rho(0)$ for two sets both with $U_1=U_2=U_{12}=0.05$,
but set 1 with $\pi\Gamma_1=\pi\Gamma_2=0.01$ and set 2 with $\pi\Gamma_1=0.01$, $\pi\Gamma_2=0.007896$.
In each  case the levels are adjusted to give an SU(4) point $n_{\rm tot}=1$. In the second set
the difference between the energy levels has to be adjusted so that the effective levels coincide. As consequence the  spectral density on two dots,
$\rho_1(\omega)$ and $\rho_2(\omega)$ differ. Nevertheless all the spectral densities remain very close
in the low energy regime near the Fermi level, indicating that the SU(4) symmetry can largely  be restored in this regime by adjusting the difference in the levels on the dots, but not on the higher energy scales. \par

\section{Dependence on on-site and inter-site interactions}

\begin{figure}[!htbp]
\begin{center}
{\includegraphics[width=0.8\linewidth]{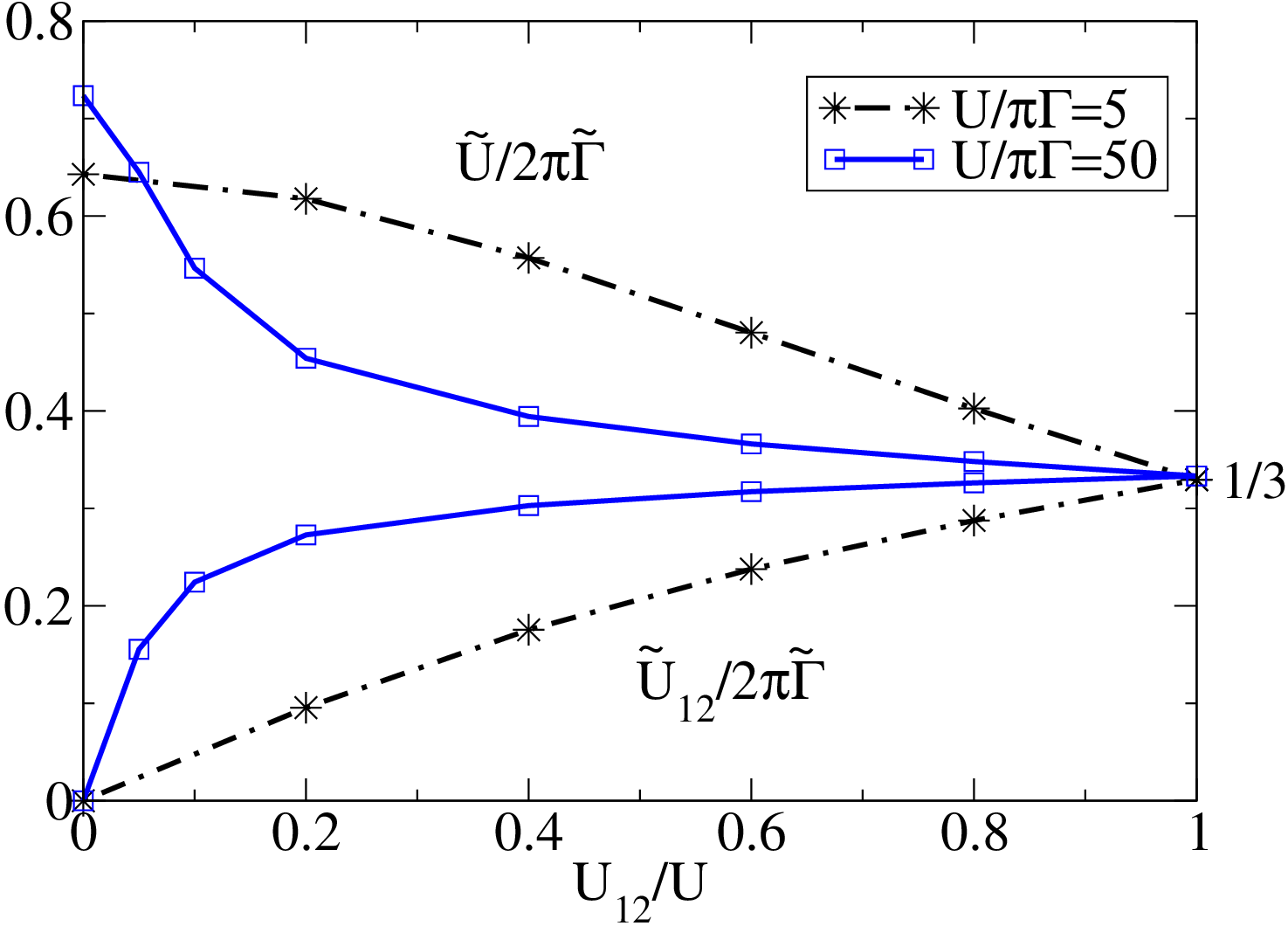}}
\caption
{(Color online) A plot of $\tilde U/2\pi\tilde\Gamma$ (higher curves) and   $\tilde U_{12}/2\pi\tilde\Gamma$ (lower curves) against the ratio $U_{12}/U$ for the  model with $n_{tot}=1$ for $U/\pi\Gamma=5,50$ and $\pi\Gamma=0.01$. }
\label{nph}
\vspace{1.0cm}
{\includegraphics[width=0.75\linewidth]{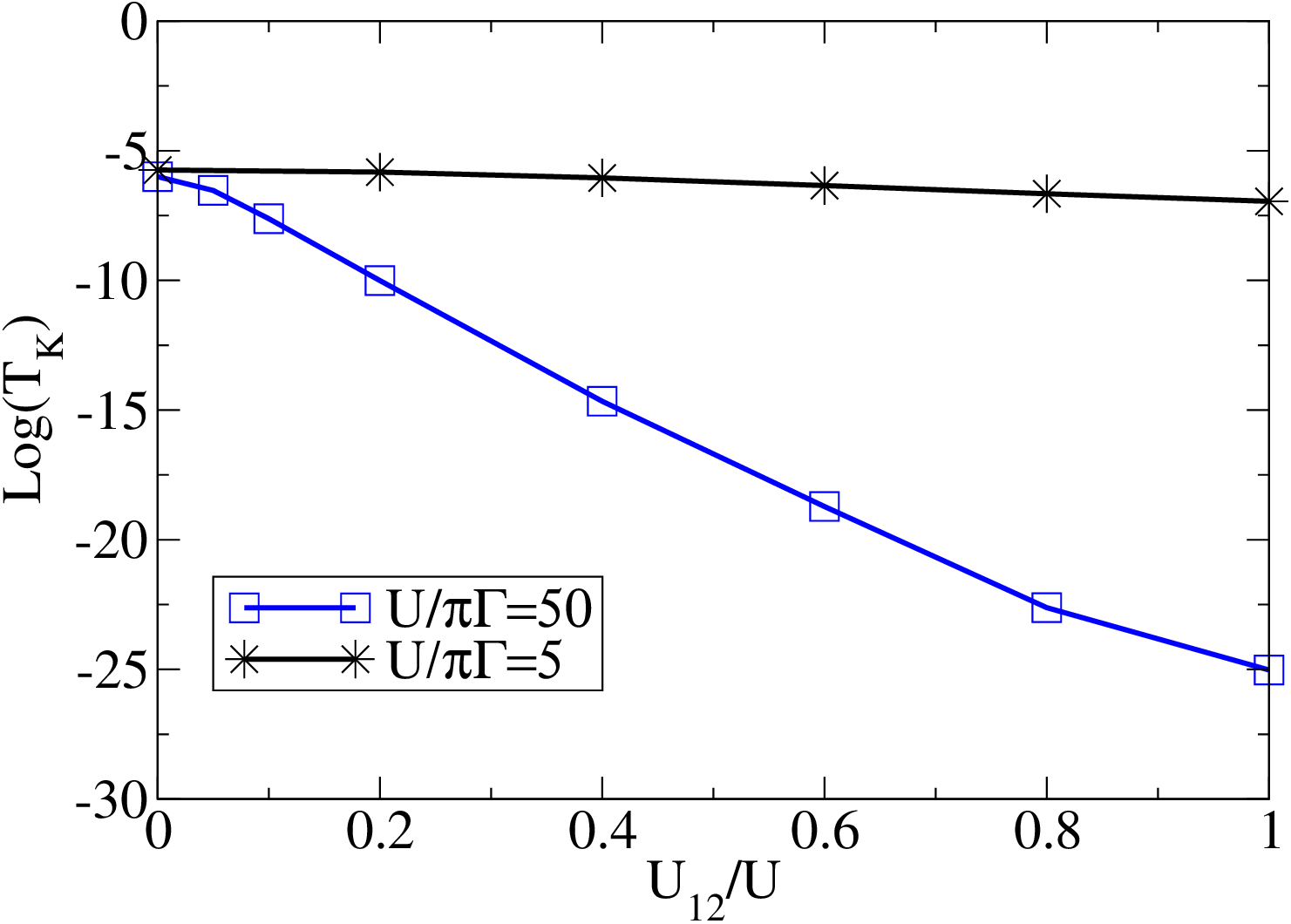}}
\caption
{(Color online)  A plot of ${\rm log}(T_{\rm K})$ versus $U_{12}/U$ for the same parameter sets as in Fig. \ref{nph}. }
\label{TK_n_1}
\end{center}
\end{figure}


We now investigate more systematically how close we can approach an SU(4) fixed point given the
interaction parameters $U_{12}$ and $U_1=U_2=U$, without the complication of different couplings
so we take $\Gamma_1=\Gamma_2$. For a given value of $U$ and $\pi\Gamma=0.01$, we calculate the
values of $\tilde U_{12}/2\pi\tilde\Gamma$ and $\tilde U/2\pi\tilde\Gamma$ as a function of $U_{12}/U$, with $\epsilon_1=\epsilon_2=\epsilon$ determined
by the constraint $n_1=n_2=1/2$. For an SU(4) fixed point we require $\tilde U_{12}/2\pi\tilde\Gamma=\tilde U/2\pi\tilde\Gamma$ and for a universal strong coupling Kondo fixed point they should both  take the value $1/3$.
In Fig. \ref{nph} we show two such plots, one for $U/\pi\Gamma=5$ and a second for $U/\pi\Gamma=50$.
For  $U/\pi\Gamma=5$ we see a steady increase of $\tilde U_{12}/2\pi\tilde\Gamma$ with $U_{12}/U$ and a steady
decrease of $\tilde U/2\pi\tilde\Gamma$, but not until $U_{12}=U$ do they become equal.    
 For  the much larger value of $U$, $U/\pi\Gamma=50$, there is an initial accelerated  increase in $\tilde U_{12}/2\pi\tilde\Gamma$ with $U_{12}/U$, mirrored by a corresponding 
decrease in $\tilde U/2\pi\tilde\Gamma$, with the two curves moving much closer. However, even with this value of $U$,
comparable with the band with $D=1$ ($U=D/2$), we do not get full SU(4) symmetry until $U_{12}=U$.  In both cases  the value
$U$ is large enough when $U_{12}=U$ to give the universal strong correlation Kondo value $1/3$. \par
\begin{figure}[!htbp]
\begin{center}
{\includegraphics[width=0.8\linewidth]{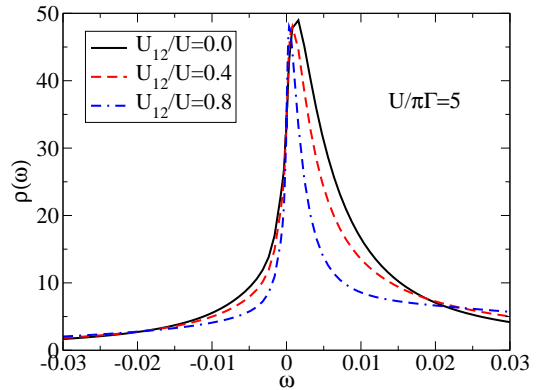}}
\caption
{(Color online) A plot of $\rho(\omega)$ against $\omega$ for ratios $U_{12}/U=0.0,0.2,0.4$ for $U=0.05$ and $\pi\Gamma=0.01$.  
 }
\label{rhos_U_0.05}
\end{center}
\end{figure}
\begin{figure}
\begin{center}
{\includegraphics[width=0.8\linewidth]{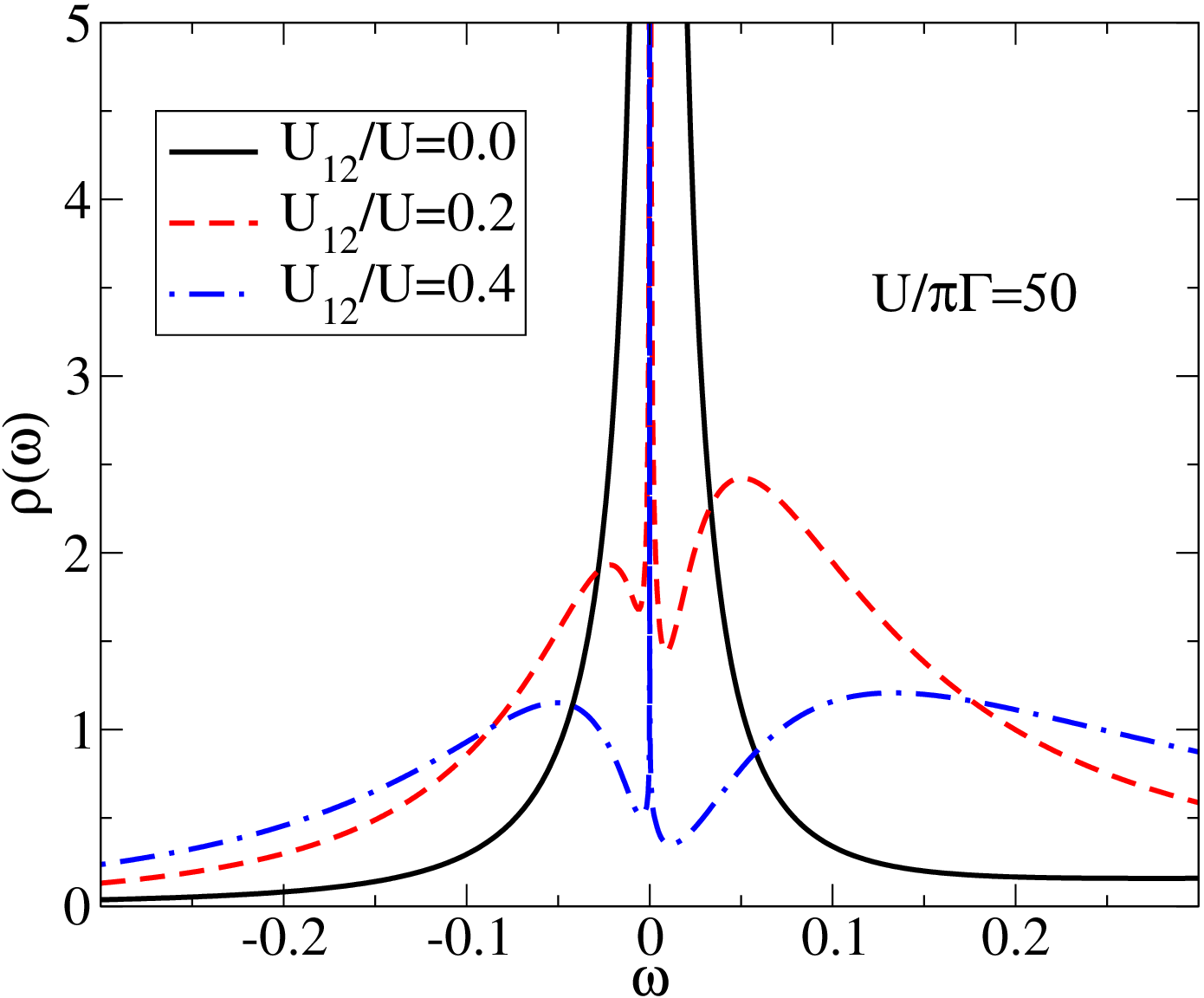}}
\caption
{(Color online) A  plot of $\rho(\omega)$ against $\omega$ for ratios $U_{12}/U=0.0,0.2,0.4$ for $U=0.5$ and $\pi\Gamma=0.01$.  
 }
\label{rhos_U_0.5}
\end{center}
\end{figure}

 There is an interesting difference in the degree of renormalization in these two cases evident
in the plot of ${\rm log}(T_{\rm K})$ shown in Fig. \ref{TK_n_1}. The Kondo temperature is a measure of the degree of renormalization as the quasiparticle weight factor $z$ is given by $z=2T_{\rm K}/\pi\Gamma$.  For $U_{12}=0$, 
there is only a modest degree of renormalization both  for $U/\pi\Gamma=5$, $z=0.641$ and
 $U/\pi\Gamma=50$, $Z=0.495$ as $U$ suppresses only the double occupation on each dot, so charge fluctuations 
between $n=0$ and $n=1$ are largely unaffected. Once $U_{12}$ is switched on these remaining charge fluctuations are also
suppressed and  the SU(4) Kondo limit, $\tilde U_{12}\tilde\rho^{(0)}(0)\to 1/3$, is approached  in both cases.
However, at this point for $U/\pi\Gamma=5$, we find $z=0.191$, being reduced by a factor of the order of 3, 
whereas for  $U/\pi\Gamma=50$, $z=2.71\times 10^{-9}$, reduced by a factor of the order $2\times 10^8$.
 The dramatic difference between the two cases can be seen in Fig. \ref{TK_n_1}  where the results for ${\rm log}(T_{\rm K})$ are plotted as a function of $U_{12}/U$.  
 This difference  in behavior in the two cases can be related to
the form of their spectral densities which are shown in Figs. \ref{rhos_U_0.05}  and \ref{rhos_U_0.5}.  In 
    Fig. \ref{rhos_U_0.05}  the spectral densities are shown for the parameter sets,  $U/\pi\Gamma=5$, for $U_{12}/U=0,0.4$ and $0.8$. There is just one
peak above the Fermi level, which does shift closer to the Fermi level and narrow as $U_{12}$ is increased. The value of
bare level parameter $\epsilon$  to give $n=1/2$ for these three cases are $\epsilon/\pi\Gamma=-0.144, -1.08, -1.82 $, all fall below the Fermi level. 
From the Friedel sum rule, we know that $n=1/2$ implies $\tilde\epsilon=\tilde\Gamma$, so the quasiparticle peak
has to lie above the Fermi level, so the peak in the spectrum is essentially that due to the renormalized quasiparticles.
It can also be interpreted as the shifted peak $\epsilon^*$ due to Haldane scaling. Haldane scaling, however, does not take into account any wavefunction renormalization, but the two shifts can be related by interpreting $\epsilon^*$ as $\epsilon+\Sigma(0)$, so that $\epsilon^*=\tilde\epsilon/z$.\par
The spectral densities for the larger $U$ case,  $U/\pi\Gamma=50$, for $U_{12}/U=0, 0.2$ and $0.4$ are shown in
Fig. \ref{rhos_U_0.5}. The corresponding values of $\epsilon$ are $\epsilon/\pi\Gamma= -0.408, -4.14, -6.80$. There is now  a very significant
change when the inter-site interaction $U_{12}$ is switched on. There is a very dramatic narrowing of the quasiparticle
peak above the Fermi level and at the same time two other peaks appear. The one below the Fermi level can be identified
as associated with the `atomic' level at $\omega=\epsilon$, and the higher peak above the Fermi level as the atomic level
at $\omega=\epsilon+U_{12}$. The picture emerging for larger $U_{12}$ begins to look similar to that for a single Anderson model near particle-hole symmetry with a three peak structure and an exponentially renormalized Kondo peak.\par

\noindent
\begin{figure}[!htbp]
\begin{center}
{\includegraphics[width=0.75\linewidth]{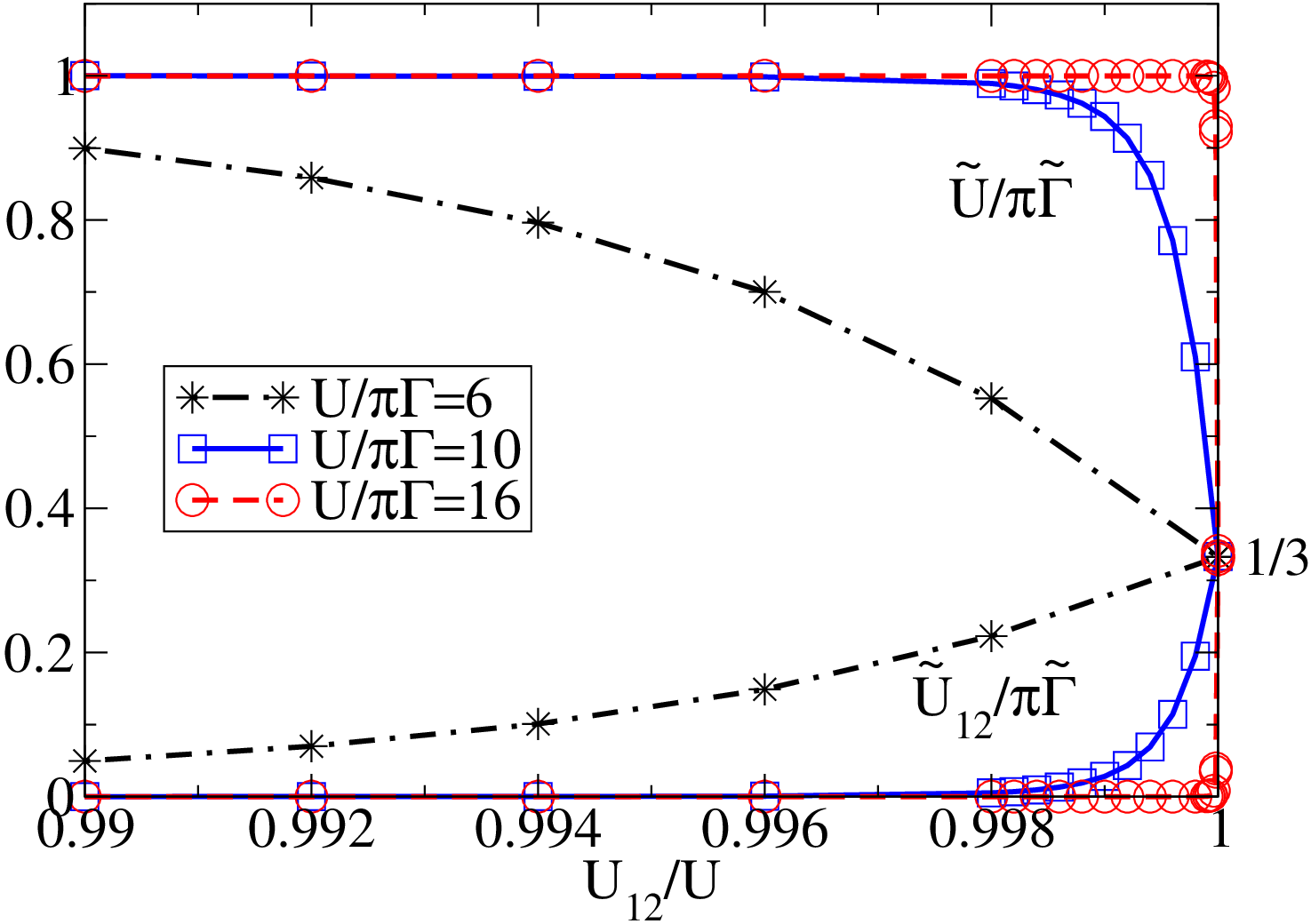}}
\caption
{(Color online)  A plot of $\tilde U/\pi\tilde\Gamma$ (higher curves) and   $\tilde U_{12}/\pi\tilde\Gamma$ (lower curves) against the ratio $U_{12}/U$ for the particle-hole symmetric model with $n_{tot}=2$ for $U/\pi\Gamma=6,10,16$ and $\pi\Gamma=0.01$. }
\label{ph}
\vspace{1.0cm}
{\includegraphics[width=0.75\linewidth]{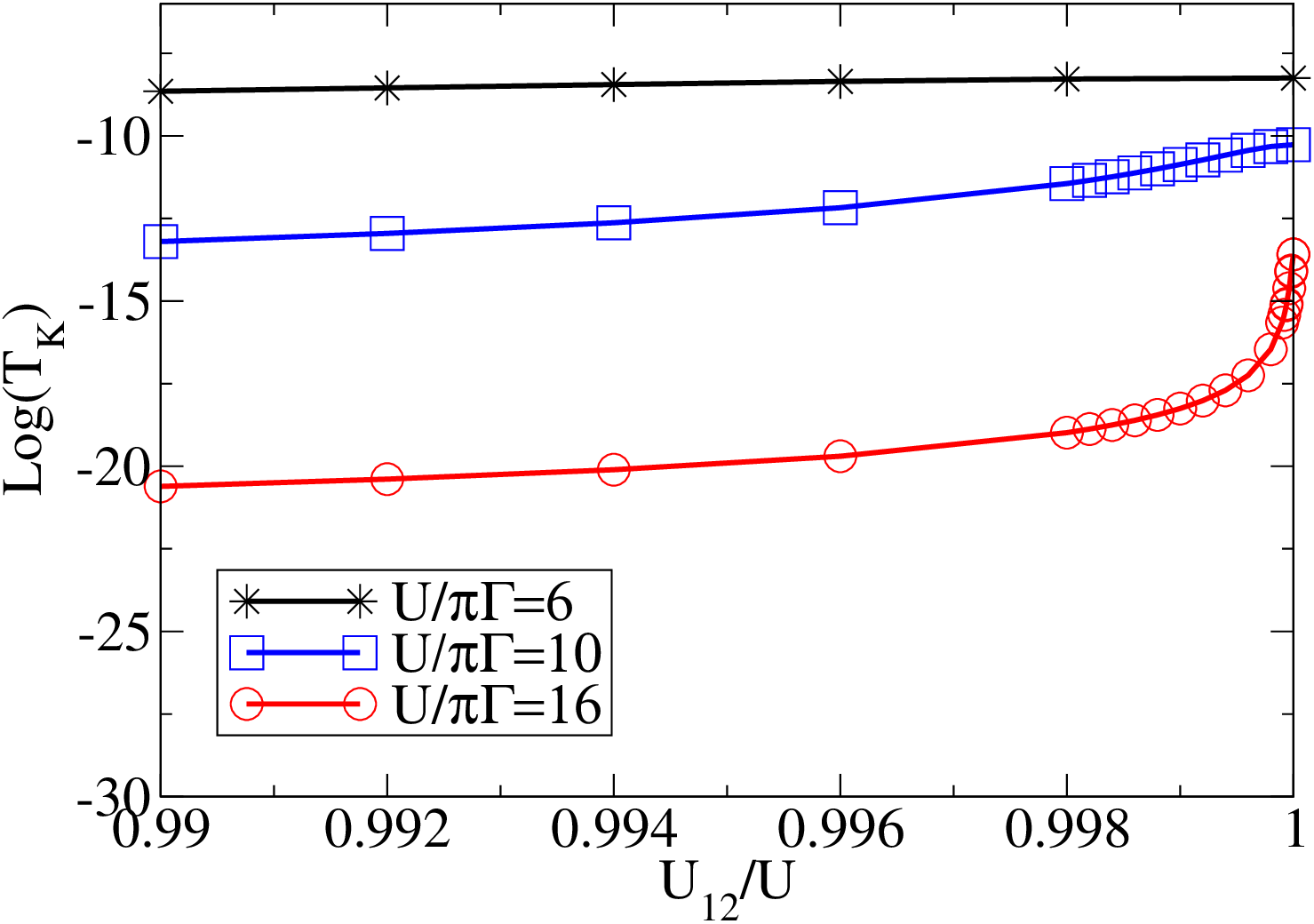}}
\caption
{(Color online) A plot of ${\rm log}(T_{\rm K})$ versus $U_{12}/U$ for the same parameter sets as in Fig. \ref{ph}. }
\label{TK_n_2}
\end{center}
\end{figure}

\noindent

It might be surprising that the condition $U_{12}/\pi\Gamma_1> 5$ is not sufficient 
to lead to an SU(4) Kondo fixed
point for $n_{\rm tot}=1$. A similar situation also occurs in the Kondo regime for the double
dot with $n_{\rm tot}=2$. For two identical dots with particle-hole symmetry, $U_{12}=U_1=U_2=U$ and $U/\pi\Gamma\gg 1$, the model maps into an SU(4) Kondo model, with the operators corresponding to
a 6-dimensional representation, rather than the 4-dimensional representation for
 $n_{\rm tot}=1$. In earlier calculations\cite{NCH12b} for $U_{12}<U$ with $U/\pi\Gamma=5$, we found only
an SU(2) fixed point as $\tilde U_{12}\sim 0$, until $U_{12}$ almost reached the value $U$,
and then a very rapid cross-over to an SU(4) fixed point as $U_{12}\to U$. It might be argued that
increasing  $U/\pi\Gamma$ will increase the range of $U_{12}/U$ favouring the SU(4) fixed point. 
However, we find the opposite is the case. In Fig. \ref{ph} (i) we plot both $\tilde U/\pi\tilde\Gamma$
and $\tilde U_{12}/\pi\tilde\Gamma$
as a function of  $ U_{12}/U$ for two
identical particle-hole symmetric dots with $U/\pi\Gamma=6,10,16$ and $\pi\Gamma=0.01$.
We see that for $U/\pi\Gamma=10$, we have a clear SU(2) fixed point with  $\tilde U/\pi\tilde\Gamma=1$
and $\tilde U_{12}/\pi\tilde\Gamma\sim 0$ over 99.8\% of the range of  $ U_{12}/U$ and an even
greater range for  $U/\pi\Gamma=16$. This reinforces the assertion that inter-dot repulsion plays a rather
different role on the lowest energy scales compared with the on-site term.\par
\begin{figure}[!htbp]
\begin{center}

\includegraphics[width=0.75\linewidth]{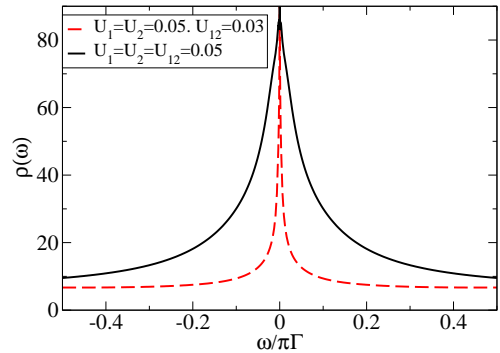}
\caption
{(Color online)  A plot of $\rho(\omega)$ as a function of $\omega/\pi\Gamma$ for a particle-hole symmetric SU(4)
model with $U_{12}=U_1=U_2=0.05$  compared with  $\rho(\omega)$  for the set  $U_{12}=0.03$, $U_1=U_2=0.05$  ($\pi\Gamma_1=\pi\Gamma_2=0.01$). }
\label{set8_12}
\end{center}
\end{figure}

In Fig. \ref{TK_n_2}  we plot the corresponding values of ${\rm log}(T_{\rm K})$ as a function of $U_{12}/U$ for the
parameter sets shown in Fig. \ref{ph}. There is very little change until $U_{12}/U>0.998$  at which point
there is an increase in $T_{\rm K}$ on the approach to the SU(4) point $U_{12}= U$, which is particularly marked
for the larger value of $U$. This is precisely opposite  behavior in the approach to the SU(4) fixed point
to that for $n_{\rm tot}=1$. In Fig. \ref{set8_12} the spectral density $\rho(\omega)$ is shown for a particle-hole symmetric case
for the sets, $U_{12}=U_1=U_2=0.05$   and $U_{12}=0.03$, $U_1=U_2=0.05$. There is a broad Kondo  peak at the Fermi level
for the first  case corresponding to an SU(4) fixed point, and an exponentially narrowed one corresponding to an SU(2) 
fixed point for the second with
a smaller value of $U_{12}$.\par

\noindent
\noindent
\section{Temperature dependence of the differential conductance}

As mentioned in the introduction the measurement of the differential conductance through a quantum dot, or arrangement of quantum dots,  has become an important way to probe locally strongly correlated states, which can be performed under equilibrium or non-equilibrium conditions.   A general formula for calculating the conductance for a single dot $i$ subject to a bias voltage $V_i$ was derived by 
Meir and Wingreen \cite{MW92}, and takes the form, 
\begin{equation}
I_i= \frac{4e\bar g_i}{\pi\hbar}\integral{\omega}{-\infty}{\infty}
[f_s(\omega)-f_d(\omega)][-{\rm Im}G^r_{i}(\omega,T, V_{ds,i})],
\end{equation}
where $\bar g_i=\Gamma_{d,i}\Gamma_{s,i}/(\Gamma_{d,i}+\Gamma_{s,i})$,
$G_{i}^r(\omega,T,V_{ds,i})$ is the steady state retarded Green's function
on the dot site, and $f_s(\omega)$, $f_d(\omega)$ are Fermi distribution
functions for the electrons in the source and drain reservoirs, respectively,
$f_{\alpha}(\omega)=f_{\rm F}(\omega-\mu_{\alpha})$ and  $\mu_{s,i}=\alpha_{s,i}eV_i$,
$\mu_{d,i}=-\alpha_{d,i}eV_i$, so that for a difference in chemical potential
across dot $i$ of $eV_i$  due to the bias voltage, $V_i$,
$\alpha_{s,i}+\alpha_{d,i}=1$. To evaluate this expression we need the retarded Green's
function as a function of the bias voltage $V_{ds,i}$. It is proving to be 
a difficult and 
challenging problem to extend the many-body techniques, such as the NRG, which can be reliably used
to tackle local strong correlation problems under equilibrium conditions, 
to non-equilibrium situations. However, the equilibrium Green's function is
sufficient to calculate the zero bias conductance, and if the coupling
of the drain to the source can be made very small, $\Gamma_{i,d}/\Gamma_{i,s}\ll 1$,
then it can be argued that the very weak current is
probing the equilibrium state of the dot.  Under these conditions useful
information can be obtained from equilibrium calculations. \par
We first of all look at the leading low temperature corrections  to  
the zero bias conductance,
\begin{equation}
G_i(T)= \frac{4e\bar
  g_i}{\hbar}\integral{\omega}{-\infty}{\infty}\beta\e^{\beta\omega}f_{\rm
  F}(\omega)^2\rho_{i}(\omega,T), 
\label{eq:GfinT}
\end{equation}
where $\beta=1/T$. 
To evaluate this expression in the low temperature regime, we use the spectral density on a given dot in terms of the renormalized parameters which is given by
 \begin{eqnarray} 
\rho(\omega,T)=z\tilde\rho(\omega,T)=\quad\quad\quad\quad\quad
\nonumber \\ \frac{1}{\pi\Gamma} \frac{\tilde\Gamma(\tilde\Gamma-\tilde\Sigma^I(\omega,T))}{(\omega-\tilde\epsilon-\tilde\Sigma^R(\omega,T))^2+(\tilde\Gamma-\tilde\Sigma^I(\omega,T))^2},
\label{sd}
\end{eqnarray}
where $\tilde\Sigma^R(\omega,T)$ and $\tilde\Sigma^I(\omega,T)$ are the real and imaginary parts of the
renormalized self-energy. For a Fermi liquid fixed point both  $\tilde\Sigma^R(\omega,T)$ and $\tilde\Sigma^I(\omega,T)$ and their first derivatives with respect to $\omega$ are zero at $\omega=0$.  The leading order temperature corrections in the Fermi liquid regime are of order $T^2$  so to evaluate the expression for $G(T)$ to this order we need the renormalized self-energy to
order, $\omega^2$ and  $T^2$. We calculate these up to second order in powers of the renormalized quasiparticle interaction
$\tilde U$ using the renormalized perturbation theory RPT\cite{Hew93}, and details of the calculation are given in the Appendix. \par
We give the result first of all for the particle-hole symmetric case $n_{tot}=2$, which is exact to this order as it
depends only on the imaginary part of the renormalized self-energy,  
 \begin{equation} 
G(T)=G(0)\left[1-(1+\phi)\frac{\pi^4}{48}\left(\frac{T}{T_{\rm K}}\right)^2+{\rm O}(T^4)\right],
\end{equation}
where  $\phi=2(\tilde U^2+2\tilde U_{12}^2)/(\pi\tilde\Gamma)^2$ is the term arising from the quasiparticle
interactions. In the SU(2) case, $\tilde U/\pi\tilde\Gamma=1$ and $\tilde U_{12}=0$, so
 $\phi^{(2)}=2$.  For SU(4) with particle-hole symmetry, $\tilde U_{12}=\tilde U$ and  $\tilde U/\pi\tilde\Gamma=1/3$, so  $\phi^{(4)}=2/3$. We note that the correction term due to the quasiparticle interactions is smaller in the SU(4) case. This is in line with results for $N$-fold degenerate Anderson and Kondo models, where the effects of the quasiparticle interactions
tend to zero in a  suitably
scaled large $N$ limit. We also note that leading term is negative so that the very low temperature conductance
decreases with temperature. 

\par
In contrast the result for the SU(4) model with $n_{\rm tot}=1$ has an initial increase with temperature and takes the form,
 \begin{equation} 
G(T)=G(0)\left[1+(1-\psi)\frac{\pi^4}{24}\left(\frac{T}{T_{\rm K}}\right)^2+{\rm O}(T^4)\right],
\end{equation}
where $\psi$ is the correction due to the quasiparticle interactions, which arises in this case  solely 
 from the real
part of the self-energy \cite{LSL07}.
 We evaluate this term to order $\tilde U^2$ in the RPT, corresponding to the diagrams shown in Fig. \ref{rpt1}. The total contribution to $\psi$ from the tadpole diagram  Fig. \ref{rpt1} (i) is   $-(\pi/2-1)(\tilde U^2+2\tilde U_{12}^2)/(4\pi\tilde\Gamma)^2$ or $-0.1903$, and  from the second order diagram in  Fig. \ref{rpt1} (ii)  $-0.2652(\tilde U^2+2\tilde U_{12}^2)/(4\pi\tilde\Gamma)^2$ or $-0.0884$.
 The net result for $\psi$ is $\psi=-0.279$. There can be higher order corrections as the second order result for the real part of the renormalized self-energy is not exact to second order. However, this result can be expected to be a reasonable
order of magnitude estimate of the corrections arising from the quasiparticle interactions. We note that in this case  there is an initial increase of $G(T)$ with $T$.\par

The difference in the behavior of $G(T)$  for the SU(4)  cases with $n_{\rm tot}=2$  and $n_{\rm tot}=1$ can
be related to the differences in their spectral densities $\rho(\omega)$ for small $\omega$. 
For $n_{\rm tot}=2$ the spectral density has a narrow Kondo peak centred at the Fermi level
so the spectral density falls off from $\omega=0$ with a negative curvature. When
 $n_{\rm tot}=1$, on the other hand, the Kondo peak is at $\omega\sim T_{\rm K}$ above the
Fermi level, so initially rises strongly from $\omega=0$ with positive curvature,
leading to an increase of conductance with temperature. There is a corresponding
contrast in low $T$ behavior in other physical properties of SU(N) models.
For example, the impurity susceptibility $\chi(T)$ of an SU(N) Kondo model\cite{Raj83} shows an
initial rise and a maximum  with  increase of $T$ for $N>3$ (though the peak is a relatively
shallow one for $N=4$) and an initial decrease
for $N=2$.
This difference can be understood in terms of the quasiparticle
density of states,
\begin{equation}
\tilde\rho(\omega)=\frac{1}{\pi}\frac{\tilde\Gamma}{(\omega-\tilde\epsilon)^2+\tilde\Gamma^2},   
\end{equation}
 and the Friedel sum rule,
\begin{equation}
n= 1-\frac{2}{\pi}{\rm tan}^{-1}\left(\frac{\tilde\epsilon}{\tilde\Gamma}\right).   
\end{equation}
The result for the sum rule can be obtained by integrating the quasiparticle density of 
states up to the Fermi level. Hence for $n=1$ ($n_{tot}=2$, half filling), the quasiparticle density 
of states has to be centred at the Fermi level, whereas for $n=2/N$, $N>2$ ($1/N$th filling), the 
quasiparticle peak
has to lie above the Fermi level\cite{Hew93b}. For  $n=1/2$ ($n_{\rm tot}=1$) and $N=4$,
the peak is a $\omega=\tilde\epsilon=\tilde\Gamma=2T_{\rm K}/\pi$, so we have an upward curvature in $\tilde\rho(\omega)$,
and a consequent initial increase of the conductance with temperature. \par

\bigskip
\par

\section{Conclusions}

 To observe a low energy SU(4) Kondo fixed point behavior in a capacitively coupled quantum dot
we have to isolate a single  electron on the double dot system, by suppressing charge fluctuations
on the individual dots and also between the dots, such that the  occupation number on each
 dot  $n_i=1/2$. From the Friedel sum rule this implies that quasiparticle density of states, specified by
a peak at $\omega=\tilde\epsilon$ with a width $\tilde\Gamma$, must be a quarter filled with
$\tilde\epsilon=\tilde\Gamma$. The general condition to suppress charge fluctuations on a single
dot (for $U_{12}=0$) is that $U\rho(0)\gtrsim 2$, so as $\rho(0)=1/2\pi\Gamma$ quarter filling, this implies
 a value of $U$ such that $U/\pi\Gamma\gtrsim 4$. On switching on an inter-dot interaction we also need
a value of $U_{12}$ such that  $U_{12}/\pi\Gamma\gtrsim 4$ to suppress  the inter-dot charge fluctuations.
What is somewhat unexpected in our results is that these two conditions are not sufficient to
generate an SU(4) symmetric fixed point with universal spin and pseudospin Wilson ratios,
$R_s=R_{ps}=4/3$. Our calculations indicate that these conditions are satisfied only asymptotically as $T_{\rm K}\to 0$.  However, the low energy behavior of a double dot system with  $U>U_{12}>4\pi\Gamma$,
will not depend significantly on satisfying the strict criteria for SU(4) symmetry.
It depends on the form of the low energy quasiparticle spectrum, which under these conditions
corresponds to a narrow resonance just above the Fermi level. The larger the value of $U_{12}$ ($<U$)
 the greater degree of renormalization and the narrower the quasiparticle resonance. 
Hence the low energy  behavior depends on two factors, the position of the quasiparticle
resonance, which is determined by the Friedel sum rule, and the degree of renormalization,
determined by the degree to which that charge fluctuations on the individual dots and between the dots
can be suppressed. \par
 
The presence of the narrow quasiparticle resonance just above the Fermi level should
be reflected in the experimentally measured temperature dependence of the linear conductance
through a given dot, $G(T)$. This should result in an initial increase of $G(T)$ and a maximum
in contrast to the monotonic decrease which occurs when the quasiparticle peak is located at
the Fermi level. This increase is not seen in the experiments\cite{KAW14}
reporting universal SU(4) temperature dependence in a capacitatively  coupled double quantum dot,
though an initial rise is evident in the NRG calculations, with which they are compared.
  It can be argued that 
an  initial rise with temperature leading to a maximum is a clearer universal  characteristic
 low temperature feature of a  SU(4) Kondo model with $n_{tot}=1$, as it shows up in several
low temperature properties, such as in the universal temperature dependent susceptibility\cite{Raj83},
$\chi(T)$ and the universal magnetic field dependent susceptibility $\chi(H)$\cite{HR83} at $T=0$.
It does not depend on having a precise SU(4) fixed point, but only on having a narrow
resonance above the Fermi level, which is a consequence of the Friedel sum rule
and the constraint $n_{\rm tot}=1$ ($n_1=n_2=1/2$). In the SU(4)  $n_{\rm tot}=1$ Kondo limit  there is no particle-hole
symmetry for finite $T_{\rm K}$ as the peak in the quasiparticle spectrum is at
 $\omega=2T_{\rm K}/\pi$.
\par

The position of the quasiparticle peak above the Fermi level also results in a much
slower fall off of $G(T)$ with temperature at higher temperatures, than in the SU(2) case
with a Kondo resonance at the Fermi level. In our earlier work\cite{NHCB13} we showed that the features
seen in the measurements\cite{KAW14} of $G(T)$ in the higher temperature range, as a function of the dot energy level $\epsilon$,
can be interpreted in terms of the temperature dependence of the renormalized parameters for
the quasiparticles. The temperature of these parameters can be estimated from the NRG calculations. \par
The other issue we have considered here is the effect of the dots having different  couplings
to their baths, $\Gamma_1\ne\Gamma_2$, which breaks the symmetry between the dots. Tosi et al.
showed that for $U=U_{12}=\infty$, the symmetry on a low energy scale could be effectively 
restored by adjustment of the energy levels on the individual dots via the applied gate voltages.
Our calculations for finite $U$ and $U_{12}$, are largely in agreement with their conclusions
but restricted to the lowest energy scale. For $U_{12}<U$, our conclusions are in line
with the case of identical dots, that there is only an approximate SU(4) fixed point 
unless both $U$ and $U_{12}$ are greater than $D$.  

\section{Acknowledgements}
\noindent
 Numerical computation in this work was partially carried out at the Yukawa Institute Computer Facility. This work has
been supported in part by the EPSRC Mathematics Platform grant
 EP/1019111/1. One of us (OC) acknowledges the support of an EPSRC DTA
 grant EP/K502856/1. YN also acknowledges the support by JSPS KAKENHI
 Grant Number 15K05181.
\par\bigskip

\section{Appendix}
We use the renormalized perturbation theory (RPT) \cite{Hew93} to estimate the low temperature corrections arising from the interaction between the quasiparticles.
The spectral density $\rho(\omega,T)$ is given by
\begin{equation}
\rho(\omega,T)=\frac{z}{\pi}\frac{\tilde\Gamma-\tilde\Sigma^I(\omega,T)}{(\omega-\tilde\epsilon-\tilde\Sigma^R(\omega,T))^2+(\tilde\Gamma-\tilde\Sigma^I(\omega,T))^2},
\end{equation}
where $\tilde\Sigma^R(\omega,T)$ and $\tilde\Sigma^I(\omega,T)$ are the real and imaginary parts of the
renormalized self-energy. To calculate the leading low frequency and low temperature corrections to $\rho(\omega,T)$ we can use the
fact that at a  Fermi liquid fixed point that
 $\tilde\Sigma_I(\omega,T)$ and  $\tilde\Sigma_R(\omega,T)$ are both of order $\omega^2$ or $T^2$
as $\omega\to 0$ and $T\to 0$, we need only include these terms to lowest order,
\begin{eqnarray}
\frac{\rho(\omega,T)}{\rho(0,0)}=1+\pi^2\omega^2(\tilde\rho^{(0)}(0,0))^2\left(\frac{3\tilde\epsilon^2}{\tilde\Gamma^2}-1\right)
\nonumber \\
+\left(1-\frac{\tilde\epsilon^2}{\tilde\Gamma^2}\right)\pi\tilde\rho^{(0)}(0,0)\tilde\Sigma_I(\omega,T)\nonumber \\
-\frac{2\pi\tilde\epsilon\tilde\rho^{(0)}(0,0)\tilde\Sigma_R(\omega,T)}{\tilde\Gamma}
+\frac{2\pi\tilde\epsilon\omega\tilde\rho^{(0)}(0,0)}{\tilde\Gamma}.
\end{eqnarray}    
In the particle-hole symmetric case, $\tilde\epsilon=0$ this simplifies to 
\begin{equation}
\frac{\rho(\omega,T)}{\rho(0,0)}=1-\pi^2\omega^2(\tilde\rho^{(0)}(0,0))^2
+\pi\tilde\rho^{(0)}(0,0)\tilde\Sigma_I(\omega,T),
\end{equation}
so to evaluate this we need only the calculation of $\tilde\Sigma_I(\omega,T)$. To order $\omega^2$ and $T^2$ it is given exactly within the second order in the renormalized perturbation theory \cite{Hew93,NCH12a},
 \begin{equation}\tilde\Sigma_I(\omega,T)=-\frac {\pi}{2}(\tilde\rho^{(0)}(0))^3(\tilde U^2+2\tilde U_{12}^2)(\omega^2+\pi^2T^2).
\label{rse}
\end{equation}

\begin{figure}[!htbp]
\begin{center}
\includegraphics[width=0.75\linewidth]{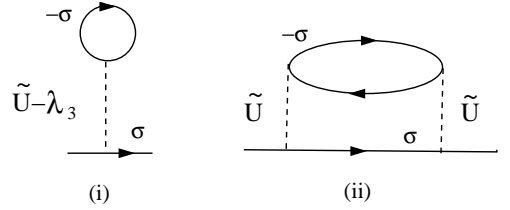}
\caption
{ The diagrams which are included in the calculation of the renormalized self-energy
to order $\tilde U^2$.}
\label{rpt1}
\end{center}
\end{figure}

\begin{figure}[!htbp]
\begin{center}
\includegraphics[width=0.75\linewidth]{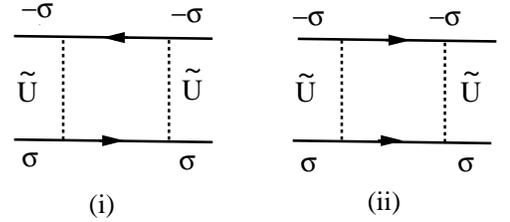}
\caption
{ The particle-hole and particle-particles scattering diagrams that contribute to the
interaction vertex counter term $\lambda_3$ 
to order $\tilde U^2$.}
\label{rpt2}
\end{center}
\end{figure}

In the SU(4) case with $\tilde\epsilon=\tilde\Gamma$, $n=1/2$ on each dot, 
\begin{eqnarray}
\frac{\rho(\omega,T)}{\rho(0,0)}=1+{2}\pi^2\omega^2(\tilde\rho^{(0)}(0,0))^2\quad\quad\nonumber \\
-2\pi\tilde\rho^{(0)}(0,0)\tilde\Sigma_R(\omega,T)+2\pi\omega\tilde\rho^{(0)}(0,0).
\end{eqnarray}
In this case the imaginary part of the renormalized self-energy makes no contribution to lowest order, but we need to evaluate the real part. We can estimate this to second order in the RPT expansion. There is a contribution to first order in $\tilde U$ and $\tilde U_{12}$ from the simple tadpole diagram (see \ref{rpt1} (i)) given by
\begin{eqnarray}
\tilde\Sigma^{(1)}_R(T)&=&\frac{\tilde U+2\tilde U_{12}}{2}(n^{(0)}(T)-n^{(0)}(0))\nonumber \\
&=&-\frac{(\tilde U+2\tilde U_{12})\tilde\epsilon(\tilde\rho^{(0)}(0,0))^2\pi^3T^2}{3\tilde\Gamma}.
\end{eqnarray} 
There is also  a second order tadpole diagram, which is essentially a first order mean field correction
to the first order tadpole diagram. The mean field equation takes the form,
\begin{equation}
\delta n(T)= 1-\frac{2}{\pi}{\rm tan}^{-1}\left(\frac{\tilde\epsilon+\tilde U\delta n(T)/2}{\tilde\Gamma}\right),
\end{equation}
where $\delta n(T)=n(T)-n(0)$. Iterating this equation to first order in $\tilde U$,
we obtain the second order correction from the second order tadpole diagram as
\begin{equation}
\frac{(\tilde U+2\tilde U_{12})^2\tilde\epsilon(\tilde\rho^{(0)}(0,0))^3\pi^3T^2}{3\tilde\Gamma}.
 \end{equation}
In the renormalized perturbation theory there are counter terms to take into account. The only counter
term we have to take explicitly into account in this second order calculation is a $\lambda_3$ counter term which is required to cancel off any zero frequency 4-vertex terms, as these have been fully included already in the definitions of $\tilde U$ and $\tilde U_{12}$. Away from particle-hole symmetry
there is a second order contribution to $\lambda_3$ given by
\begin{equation}
\lambda_3=\tilde U^2\left(\tilde\rho^{(0)}(0)-\frac{1}{\pi\tilde\epsilon}{\rm tan}^{-1}\left(\frac{\tilde\epsilon}
{\tilde\Gamma}\right)\right) ,\end{equation}
arising from the diagrams shown in Fig \ref{rpt2}. The counter term $\lambda_3$ is best handled by
carrying out the expansion in powers of $\tilde U-\lambda_3$. Hence it will give a second order contribution to the simple tadpole diagram, given by
\begin{eqnarray}
-\lambda_3 \delta n^{(0)}(T)=\frac{(\tilde U^2+2\tilde U_{12}^2)\tilde\epsilon(\tilde\rho^{(0)}(0))^2\pi^3 T^2}{3\tilde\Gamma}
\nonumber \\
\left(\tilde\rho^{(0)}(0)-\frac{1}{\pi\tilde\epsilon}{\rm tan}^{-1}\left(\frac{\tilde\epsilon}{\tilde\Gamma}\right)\right).
\end{eqnarray}
\noindent
\noindent
\noindent
\noindent

All these tadpole contributions vanish in the SU(4) case in the particle-hole symmetric limit $\tilde\epsilon\to 0$.
For the SU(4) case with $n=1/2$ on each dot with $\tilde\epsilon=\tilde\Gamma$, $\tilde U_{12}=\tilde U,$ $\tilde U\tilde\rho^{(0)}(0)=1/3$, the first two contributions from the tadpole diagram cancel
 so we are left with the contribution from the counter term only,
\begin{equation}
-\frac{(\tilde U^2+2\tilde U_{12}^2) T^2}{24\tilde\Gamma^3}\left(\frac{\pi}{2}-1\right).
\end{equation}
This gives a contribution to $\psi$,
\begin{equation}
-\frac{\pi^4 T^2}{24 T_{\rm K}^2}\left[\frac{1}{3}\left(\frac{\pi}{2} -1\right)\right],
\end{equation}
but $(\pi/2-1)/3=0.190265$.  
The remaining contribution to second order comes from the diagram in Fig. \ref{rpt1} (ii). 
 There are no counter terms to take into account explicitly. We find that this gives a contribution,
\begin{equation}
\frac{\alpha(\tilde U^2+2\tilde U_{12}^2)T^2}{24\tilde\Gamma^3},
\end{equation}
where $\alpha$ is estimated numerically as $-0.2652$.
\bibliography{artikel,biblio1}

\end{document}